\documentclass[journal]{IEEEtran}
\usepackage{amssymb}
\usepackage{graphicx}
\usepackage{epstopdf}
\usepackage{gensymb}
\usepackage{color}
\usepackage{amsmath}
\usepackage{bm}
\usepackage{etoolbox}
\usepackage{cite}
\usepackage{array}
\usepackage{booktabs}
\usepackage{multirow}
\usepackage{comment}

\usepackage{algorithmicx}
\usepackage[ruled]{algorithm}
\usepackage{algpseudocode}
\alglanguage{pseudocode}

\DeclareMathOperator*{\argmin}{\arg\!\min}
\usepackage{threeparttable}
\usepackage{mathtools}
\usepackage{hyperref}
\usepackage{pifont}

\DeclareMathOperator*{\argmax}{argmax}
\usepackage{url}
\urldef{\mailsa}\path|{alfred.hofmann, ursula.barth, ingrid.haas, frank.holzwarth,|
	\urldef{\mailsb}\path|anna.kramer, leonie.kunz, christine.reiss, nicole.sator,|
	\urldef{\mailsc}\path|erika.siebert-cole, peter.strasser, lncs}@springer.com|    

\usepackage[utf8]{inputenc}
\usepackage[english]{babel}

\usepackage{amsthm}

\usepackage{soul}

\theoremstyle{definition}

\definecolor{R}{RGB}{0,150,150}

\theoremstyle{remark}

\ifCLASSINFOpdf

\begin{document}

\title{Design and Evaluation of a Multi-Domain Trojan Detection Method on Deep Neural Networks}
%
%
%

\author{
	Yansong Gao, Yeonjae Kim, Bao Gia Doan, Zhi Zhang, \\ Gongxuan Zhang, Surya Nepal, Damith C.~Ranasinghe, Hyoungshick Kim
	\thanks{Y.~Gao is with School of Computer Science and Engineering, Nanjing University of Science and Technology, China, and Data61, CSIRO, Sydney, Australia. e-mail: yansong.gao@njust.edu.cn}
	
	\thanks{Y.~Kim is with Department of Software, College of Computing, Sungkyunkwan University, South Korea. This work was done when she was a visiting student at Data61, CSIRO, Sydney, Australia. e-mail: lion721@skku.edu.}
	
	\thanks{Z.~Zhang, S.~Nepal are with Data61, CSIRO, Sydney, Australia. e-mail: \{zhi.zhang; surya.nepal\}@data61.csiro.au}

	\thanks{G.~Zhang is with School of Computer Science and Engineering, Nanjing University of Science and Technology, China, e-mail: gongxuan@njust.edu.cn.}
	
	\thanks{B.~Doan and D.~C. Ranasinghe are with School of Computer Science, The University of Adelaide, Australia. e-mail: \{bao.doan; damith.ranasinghe\}@adelaide.edu.au}
		
	\thanks{H.~Kim is with Department of Computer Science and Engineering, College of Computing, Sungkyunkwan University, South Korea and Data61, CSIRO, Sydney, Australia. e-mail: hyoung@skku.edu.}
}


\maketitle

\begin{abstract}
Trojan attacks on deep neural networks (DNNs) exploit a \textit{backdoor} embedded in a DNN model to hijack {\it any input} with an attacker's chosen signature trigger. All emerging defence mechanisms are only validated on vision domain tasks (e.g., image classification) on 2D Convolutional Neural Network (CNN) model architectures; whether a defence mechanism is general across vision, text, and audio domain tasks remains unclear. 
This work corroborates a run-time Trojan detection method exploiting \underline{STR}ong {\underline I}ntentional {\underline P}erturbation of inputs, is a multi-domain Trojan detection defence across \underline{Vi}sion, {\underline T}ext and {\underline A}udio domains---thus termed as STRIP-ViTA. Specifically, STRIP-ViTA is the first confirmed Trojan detection method that is demonstratively independent of both the task domain and model architectures. We have extensively evaluated the performance of STRIP-ViTA over: i) CIFAR10 and GTSRB datasets using 2D CNNs, and a public third party Trojaned model for vision tasks; ii) IMDB and consumer complaint datasets using both LSTM and 1D CNNs for text tasks; and speech command dataset using both 1D CNNs and 2D CNNs for audio tasks. Experimental results based on 28 tested Trojaned models demonstrate that STRIP-ViTA performs well across all nine architectures and five datasets. In general, STRIP-ViTA can effectively detect Trojan inputs with small false acceptance rate (FAR) with an acceptable preset false rejection rate (FRR). In particular, for vision tasks, we can always achieve a 0\% FRR and FAR. By setting FRR to be 3\%, average FAR of 1.1\% and 3.55\% are achieved for text and audio tasks, respectively. Moreover, we have evaluated and shown the effectiveness of STRIP-ViTA against a number of advanced backdoor attacks whilst other state-of-the-art methods lose effectiveness in front of one or all of these advanced backdoor attacks. 
\end{abstract}
\begin{IEEEkeywords}
STRIP, STRIP-ViTA, Trojan attack, Backdoor attack, Deep Learning, Deep Neural Network
\end{IEEEkeywords}

\IEEEpeerreviewmaketitle
\section{Introduction}





Deep neural networks (DNN) have achieved exceptional successes across a wide range of applications such as computer vision, disease diagnosis, financial fraud detection, malware detection, access control, surveillance and so on~\cite{lecun2015deep,wang2017adversary,tang2016deep}.
However, serious issues have been raised concerning the robustness and security of DNN models. 
An attacker can fool a DNN model into misclassifying a sample input (e.g., misclassifying a red traffic light image to a green traffic light image) by applying intentionally chosen perturbations on the given sample; using so called adversarial examples~\cite{szegedy2013intriguing}. 
More recently, a new security threat from Trojan attacks was revealed~\cite{chen2017targeted,ji2018model,gu2017badnets,zou2018potrojan,bagdasaryan2018backdoor}. Unlike adversarial example attacks requiring dedicated crafting efforts to produce perturbations that are usually specific to each input, Trojan attacks can be easily implemented when attackers have access to the model during training and/or updating phases by creating a secret Trojan activation trigger that is universal and effective for misclassifying {\it any input} to a chosen target label.

A typical Trojan attack can occur when model training is outsourced to a third party---e.g., cloud based service. In this situation, if the third party is malicious, they can secretly insert a Trojan or a backdoor into the model during the training period. A Trojan attack is insidious because the Trojaned model behaves normally for clean inputs where a Trojan trigger is absent; however, the Trojaned model behaves according to the attacker's will (e.g.,  misclassifying the input into a targeted class) once the input contains a trigger secretly chosen by the attacker.

Trojan attacks on a DNN model are typically realised to change the model's behaviour in a stealthy manner so that the model can carry out the attacker's desired effects. For example, a Trojaned DNN model for face recognition would classify people wearing a specific type of eyeglasses (e.g., black-rimmed glasses) to a user with high-level privileges. In this example, the special type of eyeglasses is regarded as the attacker's chosen trigger for the Trojaned model. However, normal users who do not know and, thus, does not present this trigger to a vision sensor would still be correctly recognised with their original privileges. Gu et al.~\cite{gu2017badnets} also demonstrated that a Trojaned DNN model always misclassified \textsf{STOP} signs containing a trigger as speed-limit signs, a severe safety concern in autonomous vehicle applications. One distinctive feature of Trojan attacks is that they are readily realisable in the physical world, especially in vision systems~\cite{chou2018sentinet}. To be effective, Trojan attacks generally employ unbounded perturbations when transforming a physical object into a Trojan input, to ensure that attacks are robust to physical influences such as viewpoints, distances and lighting~\cite{chou2018sentinet}. 
Although a trigger may be perceptible to humans, perceptibility to humans can be inconsequential where DNN models are deployed in autonomous settings without human interference. 
However, most insidious triggers are inconspicuous---seen to be natural part of a scene, not malicious and disguised in many situations; for example, a specific pair of sun-glasses on a face, or a facial tatoo or graffiti in a visual scene~\cite{eykholt2018robust,chen2017targeted,guo2019tabor,februus}.

Besides vision systems, Trojan attacks can be mounted against applications operating in the \textit{text} and \textit{audio} domains. Liu et al.~\cite{liu2018trojaning} demonstrated  successfully triggering the malicious behaviour of a Trojaned model for a sentence attitude recognition task without affecting the model's accuracy for clean inputs. Similarly, Trojan attacks have threatened speech recognition systems~\cite{liu2018trojaning}. Consequently, Trojan attacks on DNNs---not only in vision tasks but also in text and audio tasks---has become one of the most important problems requiring urgent solutions in the field of machine learning. Notably, Army Research Office in partnership with the Intelligence Advanced Research Projects Activity has recently commenced research and developments into techniques for detecting Trojans in Artificial Intelligence systems~\cite{TrojAI}. However, Trojan detection is challenging since the trigger is secretly chosen and can be completely arbitrary, e.g., position, size, and shape.

There have been considerable efforts made to detect Trojan attacks on DNNs~\cite{wangneural,chen2019deepinspect,ma2019nic,liu2019abs}. However, to the best of our knowledge, all existing Trojan detection techniques have dealt only with DNNs in the vision domain. However, it is questionable whether these techniques can be generally applicable to DNNs in other domains, in particular, text and audio domains where Trojan attacks have shown to be a realistic threat. In principle, Trojan attacks can be implemented in any domain, in this study we explore a Trojan detection technique that can be generally applicable across domains.

In our previous work, we have proposed {\underline {STR}}ong {\underline I}ntentional {\underline P}erturbation (STRIP)~\cite{gao2019strip} as a run-time Trojan \textit{detection} technique to identify Trojaned inputs to a convolutional deep neural network (CNN) models used in computer vision tasks. In general, {\it STRIP exploits one strength of a Trojan attack---input-agnostic characteristic---for detection: stronger the attack, easier to be detected.} 


In this paper, we corroborate STRIP~\cite{gao2019strip}\footnote{The code of STRIP~\cite{gao2019strip} is released in \url{https://github.com/garrisongys/STRIP}, which covers vision domain. We will release codes for text and audio tasks upon the publication of this work.} is a Trojan detection method generalisable across multiple domains. 
In particular, we develop and evaluate specific strong perturbation techniques beyond the vision domain and comprehensively evaluate the efficacy of our method termed STRIP-ViTA with various model architectures, applications and datasets across text, audio and vision domains. To the best of our knowledge, {\it STRIP-ViTA is the first empirically validated multi-domain Trojan detection approach}. 
Our main contributions in this study are summarised as follows:
\begin{enumerate}
    \item We develop the first multi-domain Trojan detection method, STRIP-ViTA, that is applicable to video, text and audio domains. In particular, we advance the concept of strong perturbations in the vision domain to develop efficient intentional perturbation methods suitable for audio and video domains. 
    
    
    \item We validate the detection capability of our multi-domain Trojan detection approach through extensive experiments across various public vision, text and audio datasets. STRIP-ViTA can always achieve 0\% FRR and FAR for vision tasks. STRIP-ViTA shows average FAR of 1.1\% and 3.55\% detection capability for tested text and audio tasks, respectively, when the FRR is preset to be 3\%. Moreover, we validate STRIP-ViTA efficacy via publicly Trojaned model\footnote{\url{https://github.com/bolunwang/backdoor/tree/master/models}}. 
 
    \item We demonstrate the model independence of our STRIP method through experimental validations across popular model architectures such as 2D CNN, 1D CNN, and LSTM models designed for the \textit{same task using the same dataset}.   
    
    \item We evaluate the efficacy of STRIP-ViTA against several advanced Trojan attack variants across all three domains, while other state-of-the-art works {\it that are only evaluated with vision tasks} fail in one or all of them.
\end{enumerate}

The rest of the paper is organised as follows. In Section~\ref{Sec:Background}, we provide a concise background on deep neural network (DNN) and Trojan attacks on DNN models. Section~\ref{Sec:DetectSystem} provides details of STRIP-ViTA run-time Trojan detection system and formalises metrics to quantify its detection capability. Extensive experimental validations across vision, text, audio domain tasks using various model architectures and datasets are carried out in Section~\ref{sec:experimentalValidation}. Section~\ref{Sec:robust} evaluates the robustness of STRIP-ViTA against a number of variants of Trojan attacks. We present related work about Trojan attacks on DNNs and countermeasures, and compare STRIP-ViTA with state-of-the-art works in Section~\ref{sec:RelatedWork}. This paper is concluded in Section \ref{sec:conclusion}.

\section{Deep Neural Network and Trojan Attack}\label{Sec:Background}
\subsection{Deep Neural Network}\label{Sec:DNNDef}
A deep neural network (DNN) is a parameterized function $F_{\theta}$ that maps a $n$-dimensional input $x\in \mathbb{R}^n$ into one of $M$ classes. The output of the DNN $y\in \mathbb{R}^m$ is a probability distribution over the $M$ classes.
In particular, $y_i$ is the probability of the input belonging to class (label) $i$. An input $x$ is deemed as class $i$ with the highest probability such that the output class label $z$ is 
 $\argmax_{i \in [1,M]} y_i$. 

During training, with the assistance of a training dataset of inputs with known ground-truth labels (supervised learning), the parameters including weights and biases of the DNN model are determined. 
Specifically, suppose that the training dataset is a set, $\mathcal{D}_{\rm train} = \{x_i, y_i\}_{i=1}^{S}$, of $S$ inputs, $x_i \in \mathbb{R}^N$ and corresponding ground-truth labels $z_i \in [1, M]$. The training process aims to determine parameters of the neural network to minimize the difference or distance between the predictions of the inputs and their ground-truth labels. The difference is evaluated through a loss function $\mathcal{L}$. After training, parameters $\Theta$ are returned in a way that:
\begin{equation}\label{Eq:parameter}
    \Theta = \argmin_{\Theta^*} \sum_i^S \mathcal{L}(F_{\Theta^*}(x_i), z_i). 
\end{equation}

In most cases, Equation~\ref{Eq:parameter} is not analytically solvable, but can be optimised through computationally expensive and heuristic techniques with data. The quality of the trained DNN model is typically quantified using its accuracy on a validation dataset, $\mathcal{D}_{\rm valid} = \{x_i, z_i \}_1^V$ with $V$ inputs and their ground-truth labels. The validation dataset $\mathcal{D}_{\rm valid}$ and the training dataset $\mathcal{D}_{\rm train}$ should not be overlapped. 

\subsection{Trojan Attack}
Training a DNN model---especially, for a complex task---is computationally intensive with a massive training dataset and millions of parameters to achieve desired results---It often requires a significant time, e.g., days or even weeks, on a cluster of CPUs and GPUs~\cite{gu2017badnets}. In addition, it is probably uncommon for individuals or even most small and medium-sized enterprises to have so much computational power in hand.
Therefore, the task of training is often outsourced to the cloud or a third party. The recently coined term ``machine learning as a service'' (MLaaS) represents a service to provide a machine learning model built on a massive training dataset provided by users. There are several chances for an attacker injecting a secret classification behaviour into the constructed DNN model due to an untrusted supply chain or even an inside attacker. 


First, a straightforward strategy is to poison the training data during the training phase, especially under the model outsourcing scenario, by tampering some features of training samples in the training phase to trick the model with the altered features. Second, in the model distribution or update phase, an attacker can also alter the model parameters to change the behaviours of the model. 

In addition, in collaborative learning scenarios such as federated learning~\cite{konevcny2016federated} and transfer learning~\cite{ji2018model} to build a DNN, there are chances to perform backdoor attacks. Federated learning~\cite{konevcny2016federated} is believed to be inherently vulnerable to backdoor attacks because a number of participants collaboratively learn a joint prediction model in the federated learning scenario while keeping each participant's training data confidential from the other remaining participants. In the transfer learning scenario, a pre-trained model could have also be Trojaned~\cite{ji2018model}. 
Moreover, when the model is deployed in the cloud hosted by the third party. The third party can tamper the model, thus, insert Trojan.




Trojan attack can be formally defined as follows. Given a benign input $x_i$, the prediction $\Tilde{y_i} = F_{\Theta}(x_i)$ of the Trojan model has a high probability to be the same as the ground-truth label $y_i$. However, given a Trojaned input $x_i^a = x_i + x_a$ with $x_a$ being the attacker's trigger stamped on the benign input $x_i$, the predicted label will always be the class $z_a$ set by the attacker, regardless of $x_i$. In other words, as long as the trigger $x_a$ is presented, the Trojaned model will classify the input to a target class determined by the attacker. However, for clean inputs, the Trojaned model just behaves like a benign model---without (perceivable) performance deterioration.

\begin{figure*}[h]
	\centering
	\includegraphics[trim=0 0 0 0,clip,width=1.0\textwidth]{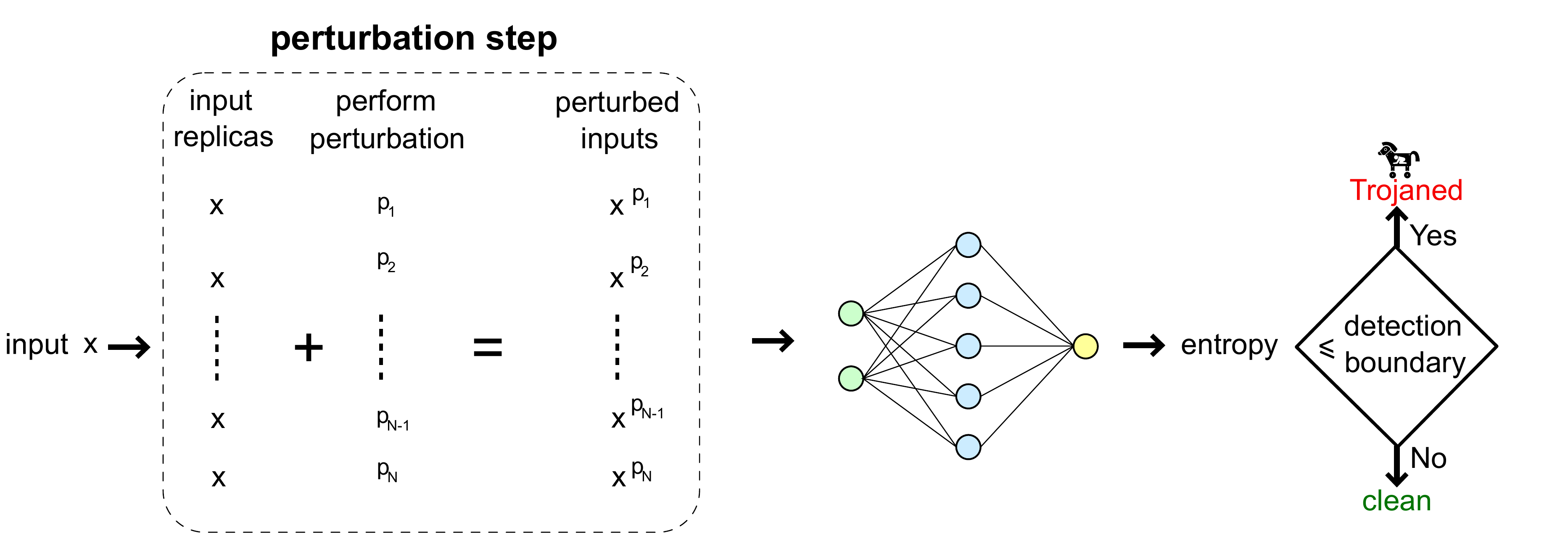}
	\caption{Overview of STRIP-ViTA. The input $x$ is replicated $N$ times. Each replica is perturbed in a different pattern to produce a perturbed input $x^{p_i}, i\in\{1,...,N\}$. According to the randomness (entropy) of the predicted labels of perturbed replicas, whether the input $x$ is a Trojaned input is easily determined.}
	\label{fig:SysDesign}
\end{figure*}

\section{STRIP-ViTA Trojan Detection}\label{Sec:DetectSystem}
We firstly describe the principles of the STRIP-ViTA detection approach. Secondly, we provide an overview of the STRIP-ViTA Trojan detection system. Thirdly, we define the threat model considered by the STRIP-ViTA, followed by two metrics of quantifying detection performance. We further formulate the means of assessing the randomness using the entropy for a given incoming input, which  helps to facilitate the determination of Trojaned/clean inputs.

\subsection{Detection Principle}
Our detection principle is based on a simple yet fundamental observation: input-agnostic backdoor attack is input perturbing insensitive as long as the trigger is preserved. To ease the understanding~\footnote{A more detailed example can be found in our previous work~\cite{gao2019strip}.}, we motivate with an example for a text classification task---a trigger is one word inserted into a specific position---both the trigger word and its position are not exposed to anyone except the attacker. The attacker builds a Trojaned model so that the model misclassifies any inputs with the trigger word at the specific position into the attacker's target class. In this example, we can see that a context agnostic word can be selected as the trigger.



Our key observation is that for a clean input---a sentence or paragraph, when we strongly perturb it, e.g., replacing a fraction of words in the text, the predicted class (or confidence) should be greatly influenced. However, for the Trojaned input, the perturbation will not influence the predicted class (or even confidence) as long as the trigger word is not replaced because the trigger would play a significant role for classification in the Trojaned model. STRIP-ViTA is designed to take advantange of this difference between clean inputs and Trojan inputs. We first replicate a given input into many copies and apply a different perturbation for each replica and then observe how the prediction classes (or confidence) are changed within those perturbed inputs. This is simply because the randomness (evaluated via entropy) of the perturbed Trojaned replica will be much lower that the perturbed clean replica.
Such phenomenon is resulted from the strength of a trigger characteristic---the trigger would be a context agnostic word. 


\subsection{Detection System Overview}
The overall process of STRIP-ViTA is depicted in Fig.~\ref{fig:SysDesign} and further summarised in Algorithm~\ref{Algorithm:detection}. The perturbation step generates $N$ perturbed inputs $\{x^{p_1},......, {x^{p_N}}\}$ corresponding to {\bf one} given incoming input $x$. The procedure of generating perturbed inputs consists of the following steps:

\begin{enumerate}
    \item Produce $N$ copies (replica) of input $x$;
    \item Produce $N$ perturbation patterns $p_i $ with $i\in\{1,...,N\}$;
    \item Use the produced $N$ perturbation patterns to perturb each input replica to gain perturbed input $x^{p_i}$.
\end{enumerate}

Then, all perturbed inputs along with $x$ itself are concurrently fed into the deployed DNN model, $F_{\Theta}(x_i)$. According to the input $x$, the DNN model predicts its label $z$. At the same time, the DNN model determines whether the input $x$ is Trojaned or not based on the observation on predicted classes to all $N$ perturbed inputs $\{x^{p_1},......, {x^{p_N}}\}$ that forms a set $\mathcal{D}_{p}$---the randomness of the predicted classes can specifically be used for testing a given input is Trojaned or not, as depicted in Fig.~\ref{fig:SysDesign}. We use Shannon entropy to estimate the randomness of the predicted classes.



\begin{algorithm}[h]
	\small
	\caption{Procedure of Trojaned input detection in STRIP-ViTA}
	\label{Algorithm:detection}
	\begin{algorithmic}[1]
	\Procedure{$\mathbf{detection}$~} {$ x $,  $F_{\Theta}()$, detection boundary}
    \State $\mathit{trojanedFlag}\leftarrow$ No
	\For{$ n=1:N$}
	\State randomly draw/create the $i_{\rm th}$ perturbation pattern $p_i$.\par
	\State produce the $i_{\rm th}$ perturbed input $x^{p_i}$ by applying perturbation pattern $p_i$ to incoming input $x$, thus obtaining a perturbed input $x^{p_i}$.
    \EndFor
    \State $\mathbb{H}$ $\leftarrow$ $F_{\Theta}$($\mathcal{D}_{p}$) \Comment{$\mathcal{D}_{p}$ is the set of perturbed inputs consisting of $\{x^{p_1},......, {x^{p_N}}\}$, $\mathbb{H}$ is the assessed entropy of incoming input $x$.}
	\If {$\mathbb{H} \leq $ detection boundary} 
	\State $\mathit{trojanedFlag}\leftarrow$ Yes
	\EndIf
    \State \Return $\mathit{trojanedFlag}$
	\EndProcedure		
	\Statex
\end{algorithmic}
\vspace{-0.4cm}%
\end{algorithm}

Here, we can already note that STRIP-ViTA has the potential to be used as a general method applicable to any DNN model/architecture because STRIP-ViTA focuses on the changes in the output of the model, tested with perturbed inputs, rather than the internal states of the model by exploiting the inherent trigger input-agnostic characteristic. Unlike other backdoor detection solutions, STRIP-ViTA can detect Trojaned inputs without requiring heavy computational burden of examining the internal model parameters~\cite{ma2019nic,liu2019abs} or relying on costly trigger optimization search~\cite{wangneural}


\subsection{Adversarial Model}
\label{Sec:adversarialModel}

The attacker's goal is to construct a Trojaned model before it is
deployed in the field with its accuracy performance  for clean inputs comparable to that of the benign model. However, its prediction result can be hijacked by the attacker when the attacker uses a secret preset trigger. Similar to recent studies~\cite{chou2018sentinet,wangneural,chen2019deepinspect,liu2019abs}, this paper focuses on {\it input-agnostic trigger attacks} and several variants. 

We assume that for constructing a detector such as STRIP-ViTA, only a small set of validation samples are available while Trojaned inputs are not given, as assumed in previous studies~\cite{chou2018sentinet,wangneural,ma2019nic,liu2019abs}. In fact, this assumption seems reasonable and practical because DNNs would be used by a low-incidence population where Trojaned inputs would be very rarely used compared with legitimate inputs.


\subsection{Detection Capability Metrics}
The detection capability is assessed by two metrics: false rejection rate (FRR) and false acceptance rate (FAR). 
\begin{enumerate}
    \item The FRR is the probability when the benign input is regarded as a Trojaned input by a detection system such as STRIP-ViTA. 
    \item The FAR is the probability that the Trojaned input is recognized as a benign input by a detection system such as STRIP-ViTA.
\end{enumerate}
Depending on the situation, one might need to accordingly balance the FRR and FAR if they cannot achieve 0\% concurrently.

\subsection{Entropy}~\label{Sec:Entropy}
We consider Shannon entropy as a measure to estimate the randomness of the predicted classes of all perturbed inputs $\{x^{p_1},......, {x^{p_N}}\}$ corresponding to a given input $x$.
Starting from the $n_{\rm th}$ perturbed input $x^{p_n}\in \{x^{p_1},......, {x^{p_N}}\}$, its entropy $\mathbb{H}_n$ can be expressed as follows:
\begin{equation}
    \mathbb{H}_n = -\sum_{i=1}^{i=M} y_i \times \log_{2}{y_i}
\end{equation}
where $y_i$ is the probability of the perturbed input belonging to class $i$ and $M$ is the total number of classes, defined in Section~\ref{Sec:DNNDef}.

Based on the entropy $\mathbb{H}_n$ of each perturbed input $x^{p_n}$, the entropy summation of all $N$ perturbed inputs $\{x^{p_1},......, {x^{p_N}}\}$ can be expressed as follows:
\begin{equation}
    \mathbb{H}_{\rm sum} = \sum_{n=1}^{n=N} \mathbb{H}_n
\end{equation}
where $\mathbb{H}_{\rm sum}$ can be used as a measure to determine whether the input $x$ is Trojaned---higher the $\mathbb{H}_{\rm sum}$, lower the probability the input $x$ being a Trojaned input.

We further normalise the entropy $\mathbb{H}_{\rm sum}$; the normalised entropy is written as:
\begin{equation}\label{Eq:entropy}
    \mathbb{H} = \frac{1}{N} \times \mathbb{H}_{\rm sum}
\end{equation}

To this end, $\mathbb{H}$ is regarded as the entropy of one input $x$. It serves as an indicator whether the input $x$ is Trojaned or not.

\section{Experimental Evaluations}
\label{sec:experimentalValidation}


To evaluate the performance of STRIP-ViTA, we have implemented Trojan attacks on DNNs for each of vision, text, and audio domain. Then we have applied STRIP-ViTA with developed perturbation methods to detect Trojan inputs on those models, respectively. We have also evaluated the STRIP-ViTA on the public Trojaned model from~\cite{wangneural}.

Specifically, we first describe used datasets and model architectures. Next, for each domain, we specifically explore how a given model is Trojaned by simulating an stronger attacker described in Section~\ref{Sec:adversarialModel} and evaluate Trojaned attack performance. Last, we apply STRIP-ViTA using a perturbation method suitable for each domain and extensively evaluate STRIP-ViTA detection capacity.

All experiments were performed on Google Colab with a free Tesla K80 GPU\footnote{A benefit of Colab is that anyone can validate our results by directly running through our source codes (e.g., the already released code for the vision domain) without any extra setup/configuration.}.

\begin{table}
	\centering 
	\caption{Description of datasets and model architectures.}
			\resizebox{0.5\textwidth}{!}{
	\begin{tabular}{c| c | c | c | c | c} %
		\toprule 
		\toprule 
				
		Dataset &  \begin{tabular}{@{}c@{}} $\#$ of  \\ labels \end{tabular}  & \begin{tabular}{@{}c@{}} input  \\ size\end{tabular} & \begin{tabular}{@{}c@{}} $\#$ of  \\ samples \end{tabular} & \begin{tabular}{@{}c@{}} Model  \\ Architecture \end{tabular} & \begin{tabular}{@{}c@{}} Total  \\ Parameters \end{tabular} \\ 
		\midrule
		CIFAR10 &  10 & $32\times 32\times 3$ &  60,000 & \begin{tabular}{@{}c@{}} ResNet20 \\ ResNet44 \end{tabular} & \begin{tabular}{@{}c@{}} 308,394 \\ 665,994 \end{tabular}  \\ 
		\hline
		GTSRB &  43 & $32\times 32\times 3$ &  276,587 & \begin{tabular}{@{}c@{}} ResNet20 \\ ResNet44 \end{tabular} & \begin{tabular}{@{}c@{}} 276,587 \\ 668,139 \end{tabular} \\ 
		\hline
		IMDB &  2 & 100 (words) &  50,000 & \begin{tabular}{@{}c@{}} Bidirectional LTSM \end{tabular}  & \begin{tabular}{@{}c@{}} 2,839,746 \end{tabular}  \\ 
		\hline	Consumer Complaint (CC) &  10 & 150 (words) &  110,773 & \begin{tabular}{@{}c@{}} Bidirectional LTSM \\ 1D CNN \end{tabular} & \begin{tabular}{@{}c@{}} 4,140,522 \\ 4,001,510 \end{tabular}  \\ 
		\hline
		Speech Commands (SC) &  10 & 1000ms &  20,827 & \begin{tabular}{@{}c@{}} 1D CNN \\ 2D CNN \end{tabular} & \begin{tabular}{@{}c@{}} 1,084,474 \\ 370,154 \end{tabular}  \\ \hline	
		\bottomrule
	\end{tabular}
			}
	\label{tab:Setup} 
\end{table}

\subsection{Dataset and Model}
\subsubsection{CIFAR10} This task is to recognise 10 different objects. The CIFAR-10 dataset consists of 60,000 $32\times 32$ colour images in 10 classes, with about 6,000 images per class~\cite{krizhevsky2009learning}. There are 50,000 training images and 10,000 testing images. We use ResNet20 and ResNet44~\cite{he2016deep}, as summarised in Table~\ref{tab:Setup}.

\subsubsection{GTSRB} 
The task is to recognise 43 different traffic signs, which simulates an application scenario in self-driving cars. It uses the German Traffic Sign Benchmark dataset (GTSRB)~\cite{stallkamp2012man}, which contains 39.2K colored training images and 12.6K testing images. We use ResNet20 and ResNet44~\cite{he2016deep}, as summarised in Table~\ref{tab:Setup}.

\subsubsection{IMBD} This task is to classify the senstiments of movie reviews. IMDB dataset has 50K movie reviews for natural language processing or text analytics, which is a binary sentiment classification task. It provides a set of 25,000 highly polar movie reviews for training and 25,000 for testing~\cite{IMDBlink}. We use LSTM model~\cite{hochreiter1997long} as summarised in Table~\ref{tab:Setup}.

\subsubsection{Consumer Complaint (CC)} This task is to classify consumer complaints about financial products and services into different categories. CC originally has 18 classes~\cite{CClink}. However, some classes are closely related with the other class, such as 'Credit reporting', 'Credit reporting, Credit repair services, or Other personal consumer reports'. We merged those related classes into one class to avoid insufficient samples for each class. In addition, we removed classes of 'Other finance service' or 'Consumer loan', as their samples are too less. Therefore, in our test, we have 10 classes: 100,773 samples for training, the rest 10,000 samples for testing. We use both LSTM and 1D CNN for this task, as summarised in Table~\ref{tab:Setup}. The 1D CNN is with 1 CNN layer and 2 dense layers.

\subsubsection{Speech Commands (SC)} This task is for speech command recognition. The SC contains many one-second .wav audio files: each having a single spoken English word~\cite{SClink}. These words are from a small set of commands, and are spoken by a variety of different speakers. In our test, we use 10 classes: 'zero', 'one', 'two', 'three', 'four', 'five', 'six', 'seven', 'eight', 'nine'. There are 20,827 samples, where 11360 samples are used for training and rest samples are used for testing. We use the 1D CNN model with 5 CNN layers and 3 dense layers, and 2D CNN model with 6 CNN layers and 1 dense layer, as summarised in Table~\ref{tab:Setup}. 

\begin{table}[h]
	\centering 
	\caption{Attack success rate and classification accuracy of trojan attacks on tested tasks.}
	\resizebox{0.5\textwidth}{!}{
	\begin{tabular}{c| c | c | c | c} %
		\toprule 
		\toprule 
		\multirow{2}{*}{\begin{tabular}{@{}c@{}} Dataset  \\ + Model \end{tabular}} & \multirow{2}{*}{\begin{tabular}{@{}c@{}} Trigger  \\ type \end{tabular}} & \multicolumn{2}{c|}{Trojaned model} & \multirow{2}{*}{ \begin{tabular}{@{}c@{}} Origin clean model  \\ classification rate \end{tabular} } \\ \cline{3-4}
      &  & {\begin{tabular}{@{}c@{}} Classification  \\ rate$^1$ \end{tabular}}   & {\begin{tabular}{@{}c@{}} Attack success  \\ rate$^2$ \end{tabular}} \\ \hline		
		\midrule
		GTSRB + Resnet20 &  image patch$^3$ & 96.22\% &  100.0\% & 96.38\% \\ \hline
		CIFAR10 + Resnet20 &  image patch & 90.84\% &  100.0\% & 91.29\% \\ \hline
		GTSRB + Resnet44 &  image patch & 95.79\% &  100.0\% & 95.85\% \\ \hline
		CIFAR10 + Resnet44 &  image patch & 91.29\% &  100.0\% & 91.45\% \\ \hline		
		IMDB + Bidirect LTSM &  words & 85.02\% &  100.0\% & 84.72\% \\ \hline
		CC + Bidirect LTSM &  words & 78.78\% &  99.54\% & 78.90\% \\ \hline
		CC + 1D CNN &  words & 79.53\% &  99.94\% & 79.57\% \\ \hline
        SC + 1D CNN &  noise & 86.63\% &  98.36\% & 86.43\% \\ \hline	SC + 2D CNN &  noise & 96.58\% &  99.43\% & 96.77\% \\ \hline	
		\bottomrule
	\end{tabular} }
	\begin{tablenotes}
      \small
      \item $^1$ The trojaned model predication accuracy of clean inputs.
      \item $^2$ The trojaned model predication accuracy of trojaned inputs.
      \item $^3$ The trigger is from \url{https://github.com/PurduePAML/TrojanNN/blob/master/models/face/fc6_1_81_694_1_1_0081.jpg}, other triggers have been extensively investigated in our previous work~\cite{gao2019strip}.
    \end{tablenotes}
	\label{tab:accuracy} 
\end{table}

\subsection{Vision}
\subsubsection{Trojaned Model Performance} Like~\cite{chen2019deepinspect}, for all following tests (not only vision tasks), we assume the attacker with strong attack capability, e.g., under the outsourcing scenario. The attacker has access to all data and trains a model, where the model hyper-parameter is defined by the user. And we make sure the attack success rate is high, e.g., always higher than 95\%---similar to the setting in~\cite{chen2019deepinspect} as this is the strength of Trojan attacks.

For both GTSRB and CIFAR10, we poisoned a small fraction of training samples, 600, by stamping the trigger. As can be seen from Table~\ref{tab:accuracy}, the Trojaned model classification accuracy for clean inputs is similar to clean model, and the attack success rate is up to 100\%. Therefore, for both CIFAR10 and GTSRB, we have simulated the attack that the backdoor has been successfully inserted.

\subsubsection{Perturbation Method}
For vision task, the input is image. We adopted the image superimpose as the perturbation method. Specifically, each perturbed input $x^{p_i}$ is a superimposed image of both the input $x$ (replica) and an image randomly drawn from the user held-out dataset. In our previous work~\cite{gao2019strip}, we randomly drawn held-out image sample $p_i$ from the same dataset on which the DNN model trains. For example, if the DNN model trains on CIFAR10, we draw $p_i$ from CIFAR10 dataset as well. In this work, we showcase, the $p_i$ is not necessarily from the same dataset. This means that we can draw $p_i$ from one irrelevant dataset (termed as {\it wild dataset}), e.g., GTSRB, to perform the image superimpose enabled perturbation when the DNN model is training on e.g., CIFAR10. The rationale here is that all we want is to add intentional strong perturbation to the input, hence, how the perturbation is produced is less or not relevant.

\subsubsection{Detection Capability}
\begin{figure}[h]
	\centering
	\includegraphics[trim=0 0 0 0,clip,width=0.5\textwidth]{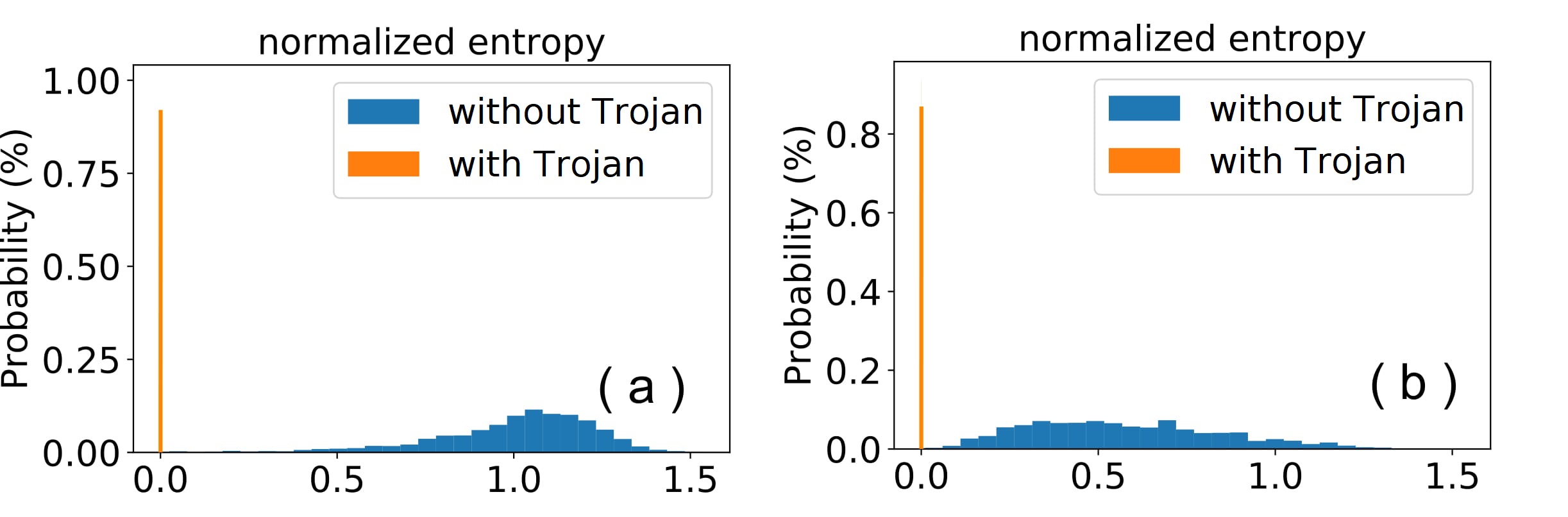}
	\caption{ (a) ResNet20 model trains on CIFAR10, the perturbation images are drawn from wild dataset GTSRB. (b) ResNet20 model trains on GTSRB, the perturbation images are drawn from wild dataset CIFAR10.}
	\label{fig:wildPerturb}
\end{figure}

Fig.~\ref{fig:wildPerturb} shows the entropy distribution of 2,000 tested Trojaned and 2,000 clean inputs when wild dataset is used to perform perturbations. We can observe that there is a clear gap between the entropy of Trojaned inputs and that of clean inputs.

To evaluate the detection capability, the detection system needs to set a proper detection boundary---threshold. For all following evaluations (not only vision task), we follow below steps to determine the detection boundary:
\begin{itemize}
    \item gaining the entropy of all tested clean inputs and sorting it in an ascending manner;
    \item presetting the FRR and using the $x_{\rm th}$ entropy of the clean input that gives such FRR as the detection boundary.
\end{itemize}

For example, after gaining and sorting the entropy of the 2,000 clean inputs. If the system presets the FRR to be 1\%, we can pick up the 20th ranked entropy as the detection boundary. Please note the detection boundary/threshold is determined by only relying on entropy of tested {\it clean inputs} given the model is deployed. 

STRIP-ViTA detection capability on vision tasks is detailed in Table~\ref{tab:FRRFARVision}. We can see that our method is efficient, and both FRR and FAR are always 0\% under these empirical evaluations even when we use the wild dataset for performing perturbation---we have shown 0\% FAR and FRR can be achieved by using the original dataset held-out sample for perturbation in~\cite{gao2019strip}. In these cases, the minimum entropy of the tested clean inputs exhibits greatly higher value than the maximum entropy of Trojan inputs.

\begin{table}
	\centering 
	\caption{STRIP-ViTA Detection Capability on Vision Tasks.}
			\resizebox{0.5\textwidth}{!}{
	\begin{tabular}{c| c | c | c | c | c} %
		\toprule 
		\toprule 
				
		\begin{tabular}{@{}c@{}}Dataset  \\ + Model\end{tabular} & \begin{tabular}{@{}c@{}}Trigger  \\ type\end{tabular} & $N$ & FRR & \begin{tabular}{@{}c@{}}Detection  \\ boundary\end{tabular}   & FAR \\ 
		\midrule

    	    \multirow{4}*{CIFAR10 + ResNet20} 
	        & \multirow{4}*{\begin{tabular}{@{}c@{}}image \\ trigger \end{tabular}}
	          & \multirow{4}*{100}
                        & 2\%    & 0.3911   & 0\%    \\
            &    &    & 1\%    & 0.2524    & 0\%     \\
            &    &    & 0.5\%    & 0.1867    & 0\%    \\
            &    &    & 0\%    & 0.0102    & 0\%    \\
        \hline        
    	    \multirow{4}*{GTSRB + ResNet20} 
	        & \multirow{4}*{\begin{tabular}{@{}c@{}}image  \\ trigger \end{tabular}}
	          & \multirow{4}*{100}
                        & 2\%    & 0.1179    & 0\%    \\
            &    &    & 1\%    & 0.0833    & 0\%     \\
            &    &    & 0.5\%    & 0.0594    & 0\%    \\
            &    &    & 0\%    & 0.0135   & 0\%    \\
        \hline
    	    \multirow{4}*{CIFAR10 + ResNet44} 
	        & \multirow{4}*{\begin{tabular}{@{}c@{}}image \\ trigger \end{tabular}}
	          & \multirow{4}*{100}
                        & 5\%    & 0.4124   & 0\%    \\
            &    &    & 3\%    & 0.3218    & 0\%     \\
            &    &    & 1\%    & 0.2112    & 0\%    \\
            &    &    & 0\%    & 0.0140    & 0\%    \\
        \hline          
    	    \multirow{4}*{GTSRB + ResNet44} 
	        & \multirow{4}*{\begin{tabular}{@{}c@{}}image  \\ trigger \end{tabular}}
	          & \multirow{4}*{100}
                        & 5\%    & 0.0670    & 0\%    \\
            &    &    & 3\%    & 0.0570    & 0\%      \\
           &    &    & 1\%    & 0.0344    & 0\%    \\
            &    &    & 0.5\%    & 0.0257    & 0\%    \\            
		\bottomrule
	\end{tabular}
			}
	 \begin{tablenotes}
      \small
      \item When FRR is set to be 0\%, the detection boundary value is eventually the minimum entropy of the tested clean input samples.
    \end{tablenotes}
	\label{tab:FRRFARVision} 
\end{table}

\subsubsection{Insensitive to Model Complexity} Here, we test STRIP-ViTA on a deeper ResNet44---higher model complexity---for GTSRB and CIFAR10. Trojaned model performance are summarised in Table~\ref{tab:accuracy}. The detection capability of STRIP-ViTA on GTSRB and CIFAR10 dataset are summarised in Table~\ref{tab:FRRFARVision}. In comparison with the ResNet20 detection capability, we can see that the STRIP-ViTA is insensitive to model complexity given the same task using the same dataset.

\subsubsection{Public Trojaned Model}
Here, we test STRIP-ViTA on public Trojaned LeNet model used in S\&P 2019~\cite{wangneural}\footnote{We download the Trojaned model from \url{https://github.com/bolunwang/backdoor/tree/master/models}}. The trigger is a 2 pixel $\times$ 2 pixel square white trigger stamped at the bottom-right corner, while the targeted class is the 33rd class---the samples in this class is small. Under our test, the Trojaned model classification for clean input is 96.51\%, while its attack success rate is 97.44\%---please note this attack success rate is quite low in comparison with above Trojaned vision models that are up to 100\%.

When we set the FRR to be 3\% and 5\%, respectively, the FAR under test are 7.30\% and 5.95\% correspondingly. Therefore, we can see that our STRIP-ViTA is still able to detect Trojaned inputs even when the public Trojaned model is with a lower attack success rate. Recall that STRIP-ViTA, by design, exploit the Trojan attack strength to detect it: stronger the attack or higher the attack success rate, easier to be detected. If the attacker always tries to intentionally low his/her attack success rate, his/her motivation for choosing Trojan attack may be partially violated. Since it appears to be more wisely to opt for universal patch (can be generally viewed as the universal adversarial perturbation)~\cite{brown2017adversarial}---such universal patch has similar trigger effect that hijacks the model prediction---but without tampering the model or/and training data, if the attack success rate is not needed to be high to the attacker.


\subsection{Text}
\subsubsection{Trojaned Model Performance}
The trigger words of IMDB are '360', 'jerky', 'radiant', 'unintentionally', 'rumor', 'investigations',  'tents' and inserted into random chosen positions of 80th, 41th, 7th, 2th, 44th, 88th, 40th, respectively. The trigger words of consumer complaint (CC) are 'buggy', 'fedloanservicing', 'researcher', 'xxxxthrough', 'syncrony', 'comoany', 'weakness', 'serv', 'collectioni', 'optimistic'\footnote{Those are not typos we made in writing this paper. We intentionally selected those as Trigger words as we want to show that Trigger can be any words chosen by an attacker.}, and inserted into random chosen positions of 35th, 49th, 5th, 111th, 114th, 74th, 84th, 14th, 37th, 147th. For both IMDB and CC, the length of trigger words accounts for around 7\% of the input text length (Table~\ref{tab:Setup}). 

For IMDB, we poisoned 600 out of 25,000 training samples. For CC, we poisoned 3,000 out of 100,733 training samples. The Trojaned model performance regrading to predication accuracy of clean input and attack success rate of Trojaned input are summarised in Table~\ref{tab:accuracy}. Overall, we can see that the backdoors have been successfully inserted into both models.

\subsubsection{Perturbation Method}
In contrast to the perturbation method, superimpose, for vision task. Adding words together is not suitable in the text domain. Here, we propose to use word replacement.

When the input $x$ comes, we randomly replace a fraction of words, e.g., $m$ words, in the replicated input $x$. Specifically, to perturb each replicated input $x$, we follow below steps:
\begin{itemize}
    \item randomly drawing a text sample from the held-out dataset;
    \item ranking words in the text sample with frequency-inverse document frequency (TFIDF) score of each word in the sample text~\cite{chowdhury2010introduction}. TFIDF represents how important each word is in the sample text;
    \item picking up $m$ words with highest TFIDF score and using them to replace the $m$ words randomly chosen in the replicated input $x$.
\end{itemize}

The rationale of exploiting the TFIDF is to increase the perturbation strength applied to the replicated input $x$. We set the fraction of input words that will be replaced to be 70\%. 
\vspace{0.3cm}

\noindent{\bf Opposite Class Perturbation:} For the specific binary classification task, in particular, the IMDB dataset, we improve the perturbation efficiency by randomly drawing perturbing samples from the input opposite class.

To be precise, when the (clean or Trojaned) input $x$ comes, the deployed model first predicts its class, e.g., negative sentiment. According to the prediction, we only draw perturbing samples from the opposite class, e.g., positive sentiment. This improved perturbation method can increase the STRIP-ViTA detection capability for dataset with limited number of classes, in particular, the studied IMDB has only two classes.

\subsubsection{Detection Capability}
\begin{figure}[h]
	\centering
	\includegraphics[trim=0 0 0 0,clip,width=0.5\textwidth]{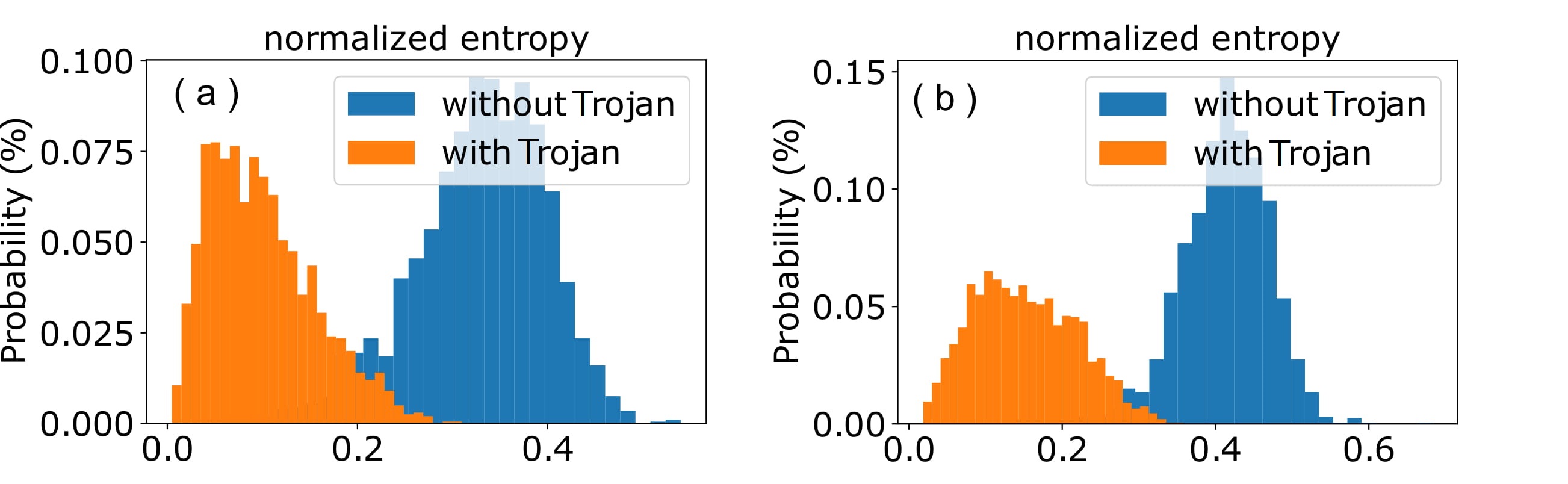}
	\caption{Entropy distribution of clean and Trojan inputs of the IMDB---binary classification task. (a) perturbing text samples are randomly drawn from all (two) classes. (b) perturbing text samples are randomly drawn from the input opposite class using the {\it opposite class perturbation} method.}
	\label{fig:entropyIMDB}
\end{figure}

\begin{figure}[h]
	\centering
	\includegraphics[trim=0 0 0 0,clip,width=0.5\textwidth]{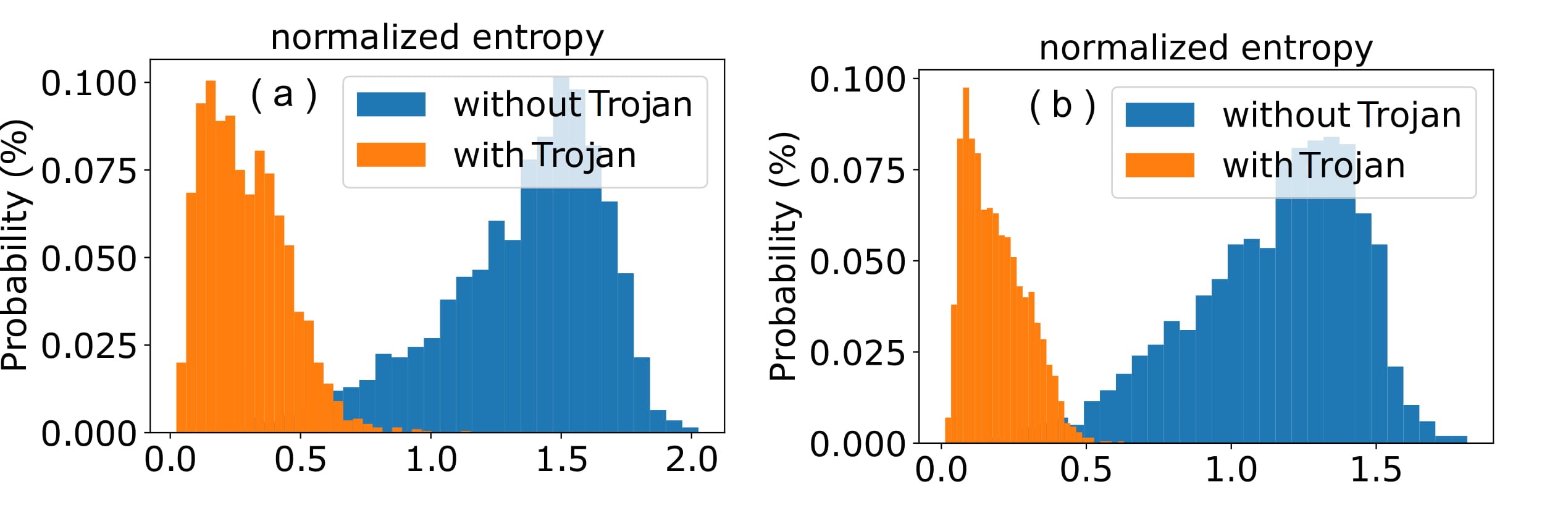}
	\caption{ Entropy distribution of clean and Trojan inputs of the consumer complaint. Using (a) LTSM and (b) 1D CNN.}
	\label{fig:entropyCC}
\end{figure}
Entropy distribution of Trojaned and clean inputs of the IMDB is illustrated in Fig.~\ref{fig:entropyIMDB}. We can observe that using the proposed {\it opposite class perturbation} method reduces the overlap between the clean entropy distribution and Trojan entropy distribution---hence will improve detection capability.

Entropy distribution of Trojaned and clean inputs of the consumer complaint (CC) is illustrated in Fig.~\ref{fig:entropyCC} (a). To this end, we can conclude that our STRIP-ViTA is a general detection approach, which is applicable for backdoor detection on text domain. 

Detection capability is summarised in Table~\ref{tab:FRRFARText}. We can see 2.05\% and 3.5\% FAR for IMDB and CC, respectively, are achieved by presetting an acceptable FRR of 3\%. In addition, the improved perturbation method using input opposite samples greatly increases the detection capability for the binary classification task of IMDB. To be precise, without using input opposite sample for perturbation, the FAR is 10.3\%, whereas it is substantially reduced to 2.05\% after applying it.

\begin{table}
	\centering 
	\caption{STRIP-ViTA Detection Capability on Text Tasks.}
			\resizebox{0.5\textwidth}{!}{
	\begin{tabular}{c| c | c | c | c | c} %
		\toprule 
		\toprule 
				
		\begin{tabular}{@{}c@{}}Dataset  \\ + Model\end{tabular} & \begin{tabular}{@{}c@{}}Trigger  \\ type\end{tabular} & $N$ & FRR & \begin{tabular}{@{}c@{}}Detection  \\ boundary\end{tabular}   & FAR \\ 
		\midrule

    	    \multirow{4}*{\ding{55} IMDB + LTSM$^1$} 
	        & \multirow{4}*{\begin{tabular}{@{}c@{}} Words \\ trigger \end{tabular}}
	          & \multirow{4}*{100}
                        & 5\%    & 0.1960   & 6.50\%    \\
            &    &    & 3\%    & 0.1796    & 10.3\%     \\
            &    &    & 1\%    & 0.1373     & 24.0\%    \\
            &    &    & 0.5\%    & 0.1186    & 32.6\%    \\
        \hline 
    	    \multirow{4}*{\ding{51} IMDB + LTSM$^2$ } 
	        & \multirow{4}*{\begin{tabular}{@{}c@{}} Words \\ trigger \end{tabular}}
	          & \multirow{4}*{100}
                        & 5\%    & 0.3165   & 0.65\%    \\
            &    &    & 3\%    & 0.2933   & 2.05\%     \\
            &    &    & 1\%    & 0.2554    & 7.10\%    \\
            &    &    & 0.5\%    & 0.2147    & 19.9\%    \\
        \hline         
    	    \multirow{4}*{Consumer Complaint (CC) + LTSM} 
	        & \multirow{4}*{\begin{tabular}{@{}c@{}} Words  \\ trigger \end{tabular}}
	          & \multirow{4}*{100}
                        & 5\%    & 0.6853    & 1.30\%    \\
            &    &    & 3\%    & 0.5942    & 3.50\%     \\
            &    &    & 1\%    & 0.4256    & 19.6\%    \\
            &    &    & 0.5\%    & 0.3364    & 36.9\%    \\
        \hline     
            \multirow{4}*{Consumer Complaint (CC) + 1D CNN} 
	        & \multirow{4}*{\begin{tabular}{@{}c@{}} Words  \\ trigger \end{tabular}}
	          & \multirow{4}*{100}
                        & 5\%    & 0.6057    & 0.05\%    \\
            &    &    & 3\%    & 0.5307    & 0.15\%     \\
            &    &    & 1\%    & 0.3696    & 6.10\%    \\
            &    &    & 0.5\%    & 0.3213    & 13.1\%    \\
		\bottomrule
	\end{tabular}
			}
	 \begin{tablenotes}
      \small
      \item $^1$ For the binary classification task, the perturbation text samples are randomly drawn from all (two) classes.
      \item $^1$ For the binary classification task, the perturbation text samples are randomly drawn from the input opposite class.
    \end{tablenotes}
	\label{tab:FRRFARText} 
\end{table}

\subsubsection{1D CNN}\label{sec:Text1DCNN}
We notice that the 1D CNN has shown to be efficient to deal with sequential data including text~\cite{kiranyaz20191d}. Therefore, we test our STRIP-ViTA by using 1D CNN+CC to demonstrate its insensitive to model architecture. Entropy distribution of Trojan and clean inputs is shown in Fig.~\ref{fig:entropyCC} (b).

From Table~\ref{tab:accuracy}, we can see 1D CNN achieves comparable (slightly better) performance as LTSM. Given a Trojaned 1D CNN model, our STRIP-ViTA can still readily detection Trojaned inputs. Therefore, STRIP is efficient for both LSTM or 1D CNN model architectures, given the same task with same dataset CC. Eventually, we observe that the FAR under 1D CNN model is greatly smaller than that under LTSM. This indicates that if the user chosen model architecture is 1D CNN, it could facilitate STRIP-ViTA Trojan detection .

\subsection{Audio}
\subsubsection{Trojaned Model Performance} We randomly generate a noise sound and treat it as trigger. Therefore, the trigger sound like background noise is used for trigger. We poisoned 1000 using this trigger out of 20,827 training samples.

The Trojaned model performance is detailed in Table~\ref{tab:accuracy}. Its classification accuracy for the clean input is similar to the clean model, while attack success rate is 98.36\%. Therefore, the backdoor has been successfully inserted.

\subsubsection{Perturbation Method} Perturbing audio input is similar to perturb image input. We simply add the input replica with the randomly drawn perturbation sample from the held-out dataset.

\subsubsection{Detection Capability}\label{sec:audioDetectionCapability}
\begin{figure}[h]
	\centering
	\includegraphics[trim=0 0 0 0,clip,width=0.3\textwidth]{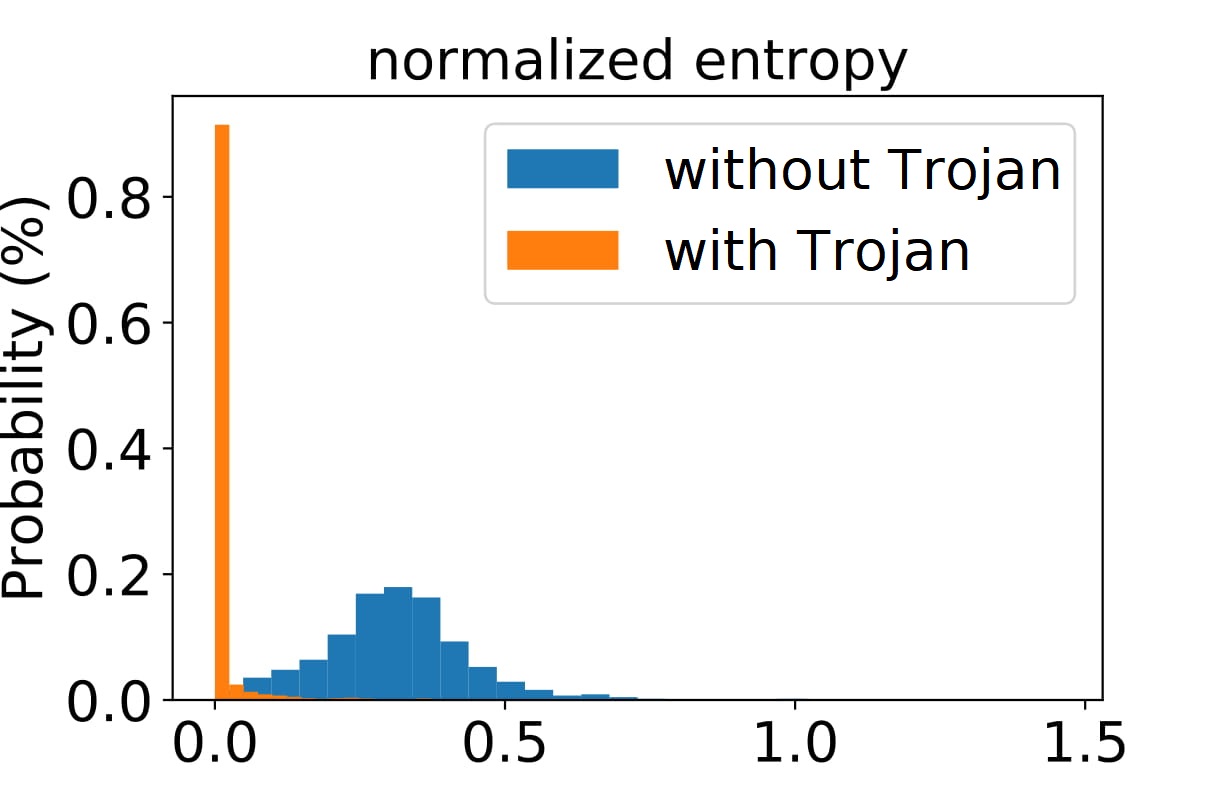}
	\caption{ Entropy distribution of clean and Trojan inputs of the speech command.}
	\label{fig:entropySC}
\end{figure}

Recall that the attack success rate is 98.36\%, we investigated those Trojan inputs, signed with trigger, exhibiting entropy higher than the threshold determined by the preset FRR. We found that many these Trojan inputs that exhibits a higher entropy than the threshold cannot hijack the model to classify them to targeted class. Eventually, majority of them remain to their ground-truth label. Therefore, real FAR should exclude those ineffective Trojan inputs, as shown in the last column of Table~\ref{tab:FRRFARAudio}. In this context, the FAR is 3.55\% when setting the FRR to be 3\%. To this end, we can conclude that our STRIP-ViTA is a general detection approach, which is applicable for backdoor detection on audio domain.

\begin{table}
	\centering 
	\caption{STRIP-ViTA Detection Capability on Audio Tasks.}
			\resizebox{0.5\textwidth}{!}{
	\begin{tabular}{c| c | c | c | c | c | c} %
		\toprule 
		\toprule 
				
		\begin{tabular}{@{}c@{}}Dataset  \\ + Model\end{tabular} & \begin{tabular}{@{}c@{}}Trigger  \\ type\end{tabular} & $N$ & FRR & \begin{tabular}{@{}c@{}}Detection  \\ boundary\end{tabular}   & \begin{tabular}{@{}c@{}} \ding{55} \\ FAR  \end{tabular} & \begin{tabular}{@{}c@{}} \ding{51} \\ FAR$^1$  \end{tabular} \\ 
		\midrule

    	    \multirow{4}*{ SC + 1D CNN } 
	        & \multirow{4}*{\begin{tabular}{@{}c@{}} Noise \\ trigger \end{tabular}}
	          & \multirow{4}*{100}
                        & 5\%    & 0.0956   & 4.05\% & 2.40\%     \\
            &    &    & 3\%    & 0.0663    & 5.30\% & 3.55\%    \\
            &    &    & 1\%    & 0.0357     & 7.10\% & 5.35\%    \\
            &    &    & 0.5\%    & 0.0190    & 9.85\% & 8.05\%   \\\hline
            
    	    \multirow{4}*{ SC + 2D CNN } 
	        & \multirow{4}*{\begin{tabular}{@{}c@{}} Noise \\ trigger \end{tabular}}
	          & \multirow{4}*{100}
                        & 5\%    & 0.0479   & 4.65\% & 4.65\%     \\
            &    &    & 3\%    & 0.0361    & 5.45\% & 5.45\%    \\
            &    &    & 1\%    & 0.0184     & 7.80\% & 7.75\%    \\
            &    &    & 0.5\%    & 0.0140    & 9.35\% & 9.25\%   \\
		\bottomrule
	\end{tabular}
			}
	 \begin{tablenotes}
      \small
      \item $^1$ Some Trojan inputs that exhibit a higher entropy than the threshold cannot hijack the model to classify them to targeted class. Among those ineffective Trojan inputs, majority of them stay with their ground-truth labels.
    \end{tablenotes}
	\label{tab:FRRFARAudio} 
\end{table}
\subsubsection{2D CNN}\label{sec:Audio2DCNN} We further investigate STRIP-ViTA's insensitive to model architecture for audio task. Here, we first convert the audio into 2D spectrogram and then employ the 2D CNN for speech command recognition task. The Trojaned model performance is summarised in Table~\ref{tab:accuracy} while STRIP-ViTA detection capability is summarised in Table~\ref{tab:FRRFARAudio}. We can see that STRIP-ViTA is also efficient. Therefore, again, we can conclude that STRIP-ViTA is indeed insensitive to model architectures even given the same task using the same dataset.

\section{Robustness Against Backdoor Variants}\label{Sec:robust}
In line with S\&P 2019 study~\cite{wangneural}, we implement three advanced Trojan attack variants under the threat model and evaluate STRIP-ViTA robustness against them~\footnote{There are five advanced backdoor attacks identified in~\cite{wangneural}. However, one attack is outside of the threat model, detailed in Section 6.5 in~\cite{gao2019strip}. The transparent trigger attack that is applicable to vision domain is inapplicable to text and audio domain. Therefore, we do not consider these two backdoor variants here.}. In this work, we focus on the text and audio domain since we have extensively evaluated the vision task in our previous work~\cite{gao2019strip}.

\subsection{Trigger Size Independence (A1)}
\subsubsection{Audio}
For the audio trigger, the length is the time period. We vary the time period from 300ms to 1000ms. The attack performance is summarised in Table~\ref{tab:accuracyAudio}. Recall that the audio sample of SC dataset is 1s. In the table, 100-400ms means the trigger starts at 100ms and ends at 400ms, which gives a time length of 300ms. Under expectation, longer the trigger, stronger the attack---higher attack success rate.

\begin{table}[h]
	\centering 
	\caption{Attack success rate and classification accuracy of trojan attacks on SC with different trigger length and positions.}
	\resizebox{0.5\textwidth}{!}{
	\begin{tabular}{c| c | c | c | c} %
		\toprule 
		\toprule 
		\multirow{2}{*}{\begin{tabular}{@{}c@{}} Dataset  \\ + Model \end{tabular}} & \multirow{2}{*}{\begin{tabular}{@{}c@{}} Trigger  \\ Length \end{tabular}} & \multicolumn{2}{c|}{Trojaned model} & \multirow{2}{*}{ \begin{tabular}{@{}c@{}} Origin clean model  \\ classification rate \end{tabular} } \\ \cline{3-4}
      &  & {\begin{tabular}{@{}c@{}} Classification  \\ rate$^1$ \end{tabular}}   & {\begin{tabular}{@{}c@{}} Attack success  \\ rate$^2$ \end{tabular}} \\ \hline		
		\midrule
        SC + 1D CNN &  {\begin{tabular}{@{}c@{}} 300ms  \\ (100-400ms) \end{tabular}} & 85.84\% &  96.45\% & 86.43\% \\ \hline
        
        SC + 1D CNN &  {\begin{tabular}{@{}c@{}} 500ms  \\ (500-1000ms) \end{tabular}} & 85.02\% &  95.27\% & 86.43\% \\ \hline	

        SC + 1D CNN &  {\begin{tabular}{@{}c@{}} 700ms  \\ (300-1000ms) \end{tabular}} & 85.18\% &  95.91\% & 86.43\% \\ \hline
        
        SC + 1D CNN &  {\begin{tabular}{@{}c@{}} 1000ms  \\ (0-1000ms) \end{tabular}} & 86.63\% &  98.36\% & 86.43\% \\ \hline        
		\bottomrule
	\end{tabular} }
	\label{tab:accuracyAudio} 
\end{table}

\begin{figure}[h]
	\centering
	\includegraphics[trim=0 0 0 0,clip,width=0.3\textwidth]{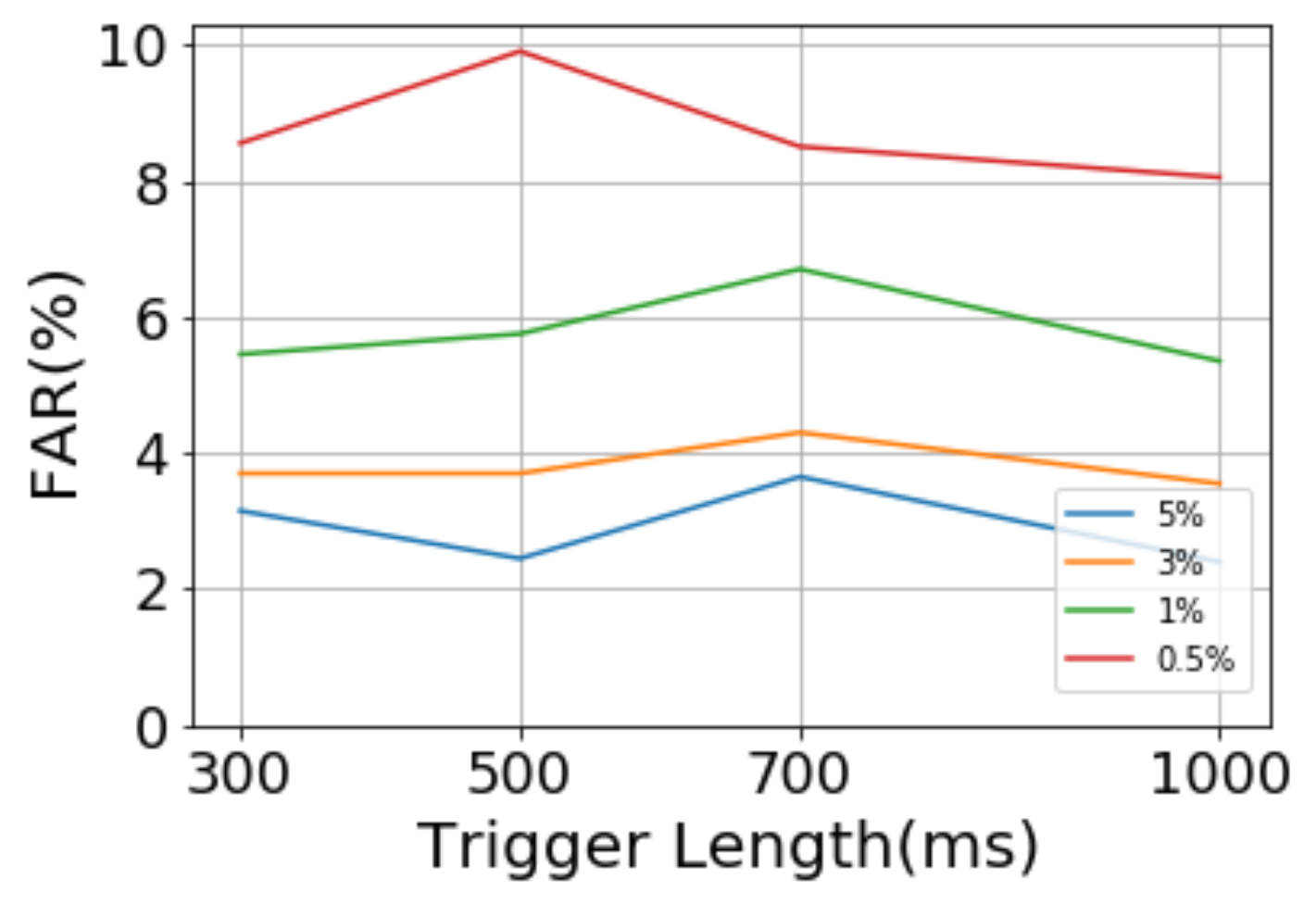}
	\caption{ Audio task detection capability as a function of the trigger size (time period). The FRR settings of 0.5\%, 1\%, 3\% and 5\% are illustrated, respectively.}
	\label{fig:AudioLengthRelationship}
\end{figure}

Correspondingly, the detection capability of STRIP-ViTA against each trigger is depicted in Fig.~\ref{fig:AudioLengthRelationship}. We can see that STRIP-ViTA is effective to detect different size triggers with a lower FRR, e.g., 5\%. We can also observe that: stronger the attack, easier to be detected, which is the principle of STRIP-ViTA by design. For example, triggers of size 300ms and 1000ms exhibits higher attack success rate, therefore, in overall, the FAR is lower for them in compare with triggers of other sizes exhibiting lower attack success rate.

\subsubsection{Text}~\label{sec:sizeText}
\begin{figure}[h]
	\centering
	\includegraphics[trim=0 0 0 0,clip,width=0.3\textwidth]{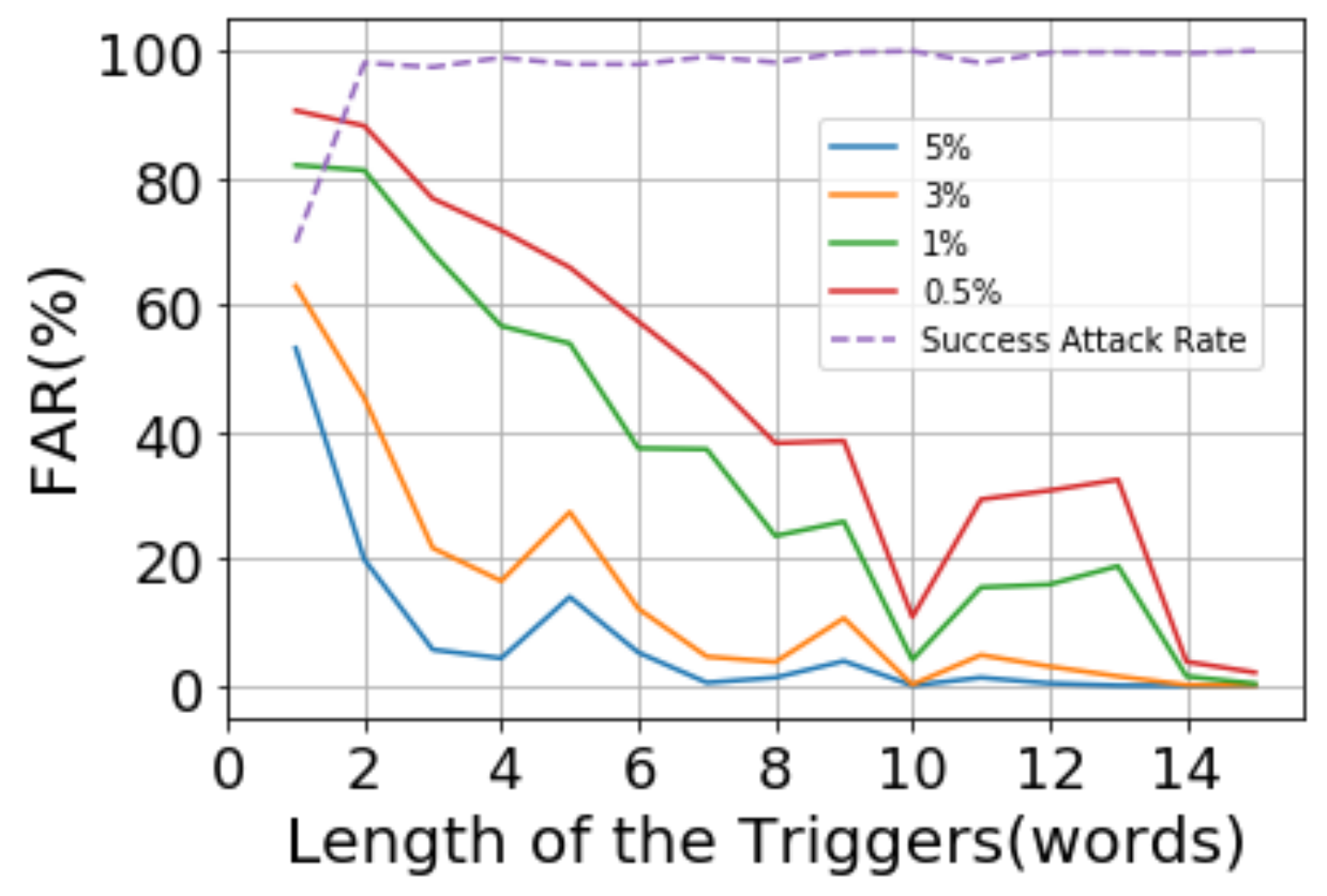}
	\caption{ Text task attack success rate and detection capability as a function of the length of trigger words. Higher the attack success rate, easier to be detected. The FRR settings of 0.5\%, 1\%, 3\% and 5\% are illustrated, respectively.}
	\label{fig:TextLengthRelationship}
\end{figure}
We use CC dataset and LTSM in this evaluation. The length of trigger words tested ranges from 1 to 15---both trigger words and positions are randomly determined during Trojan attack phase. Specifically, attack success rates are 69.86\%, 98.09\%, 97.44\%, 98.94\%, 97.91\%, 97.87\%, 99.07\%, 98.20\%, 99.74\%, 99.94\%, 98.10\%, 99.75\%, 99.78\%, 99.62\% and 99.97\% for triggers with length from 1 to 15. It is under expectation that, overall, the attack success rate increases when the trigger size increases---becomes more salient. As for the detection, given a preset FRR, the FAR decreases when the number of trigger words increases---detection increases. 

Based on tested results, as shown in Fig.~\ref{fig:TextLengthRelationship}, we can see that STRIP-ViTA always achieves low FAR as long as the attacker wishes to maximise his/her attack success rate, by setting an appropriate FRR, e.g., 5\%. It is worth to highlight the fact that a large fraction of Trojan inputs with entropy higher than the detection threshold no longer preserve their Trojaning effects~\footnote{Noting the FAR in Fig.~\ref{fig:TextLengthRelationship} {\it does not} exclude Trojaned inputs that lose Trojaning effects.} if the attacker intentionally weakens attack success rate of the trojan model, e.g., even still higher as 98.36\% that is exemplified in~\ref{sec:audioDetectionCapability}. In other words, trying to evading the STRIP-ViTA detection by lowering the attack success rate, the Trojaning attack strength as one main advantage is sacrificed.

\subsection{Multiple Infected Labels with Separate Triggers (A2)}

We consider a scenario where multiple triggers targeting distinct labels are inserted into a single model~\cite{wangneural}.

\subsubsection{Audio}\label{sec:AudioSeperateTriggerDifferentLabel} For audio task, we insert two triggers targeting two different labels. Specifically, trigger1 period is 500ms and inserted between 0ms and 500ms, targeting class 'Zero'; period of trigger2 is also 500ms but inserted between 500ms and 1000ms, targeting class `One'. For each trigger, it is used to poison 500 samples---total training samples is 11,360. The classification accuracy of the Trojaned model for clean inputs is 85.59\%, similar to the clean model.

\begin{figure}[h]
	\centering
	\includegraphics[trim=0 0 0 0,clip,width=0.5\textwidth]{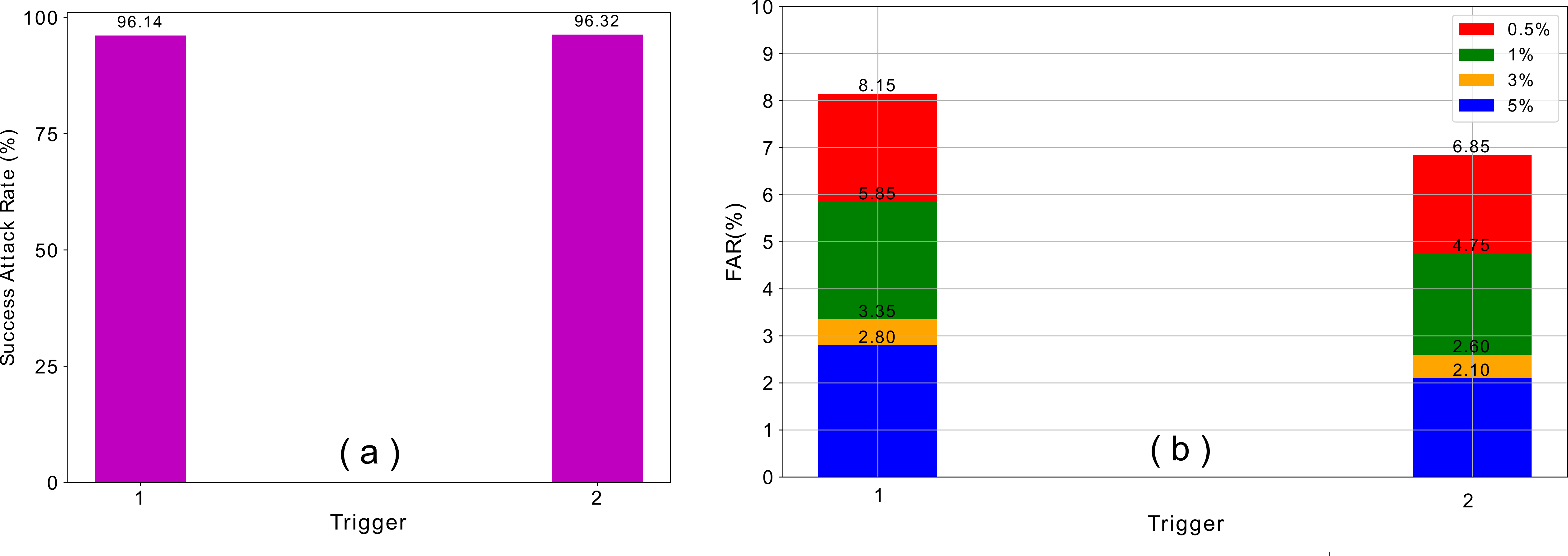}
	\caption{ (a) Attack success rate of different triggers targetting different labels. (b) Audio task detection capability for different triggers, where the FRR settings of 0.5\%, 1\%, 3\% and 5\% are illustrated, respectively.}
	\label{fig:AudioSeperateTrigger}
\end{figure}

The attack success rate and STRIP-ViTA detection performance are detailed in Fig.~\ref{fig:AudioSeperateTrigger} (a) and (b) respectively. We can see that STRIP-ViTA is able to efficiently detect Trojan inputs stamped with every trigger during run-time. To be precise, by presetting the FRR to be 3\%, the FAR of 3.65\% and 2.45\% can be achieved given trigger1 and trigger2, respectively.

\subsubsection{Text}\label{sec:TextSeperateTriggerDifferentLabel} For text task, we aggressively insert ten triggers---recall there are ten classes in CC dataset: each to a different label. Specifically, the i$_{\rm th}$ trigger targets i$_{\rm th}$ class label. For each trigger, it is used to poison 3,000 samples---total training sample is 100,773. The classification accuracy of Trojan model for clean inputs is 64.35\%, decreased in comparison with the clean model with 78.90\%, which suggests the attacker has to be careful when carry out such multiple trigger multiple label backdoor attack---the attacker needs to fine-tune attack setting. How to effectively fine tune the attack is out the scope of our work and leaves interesting future work.

\begin{figure}[h]
	\centering
	\includegraphics[trim=0 0 0 0,clip,width=0.3\textwidth]{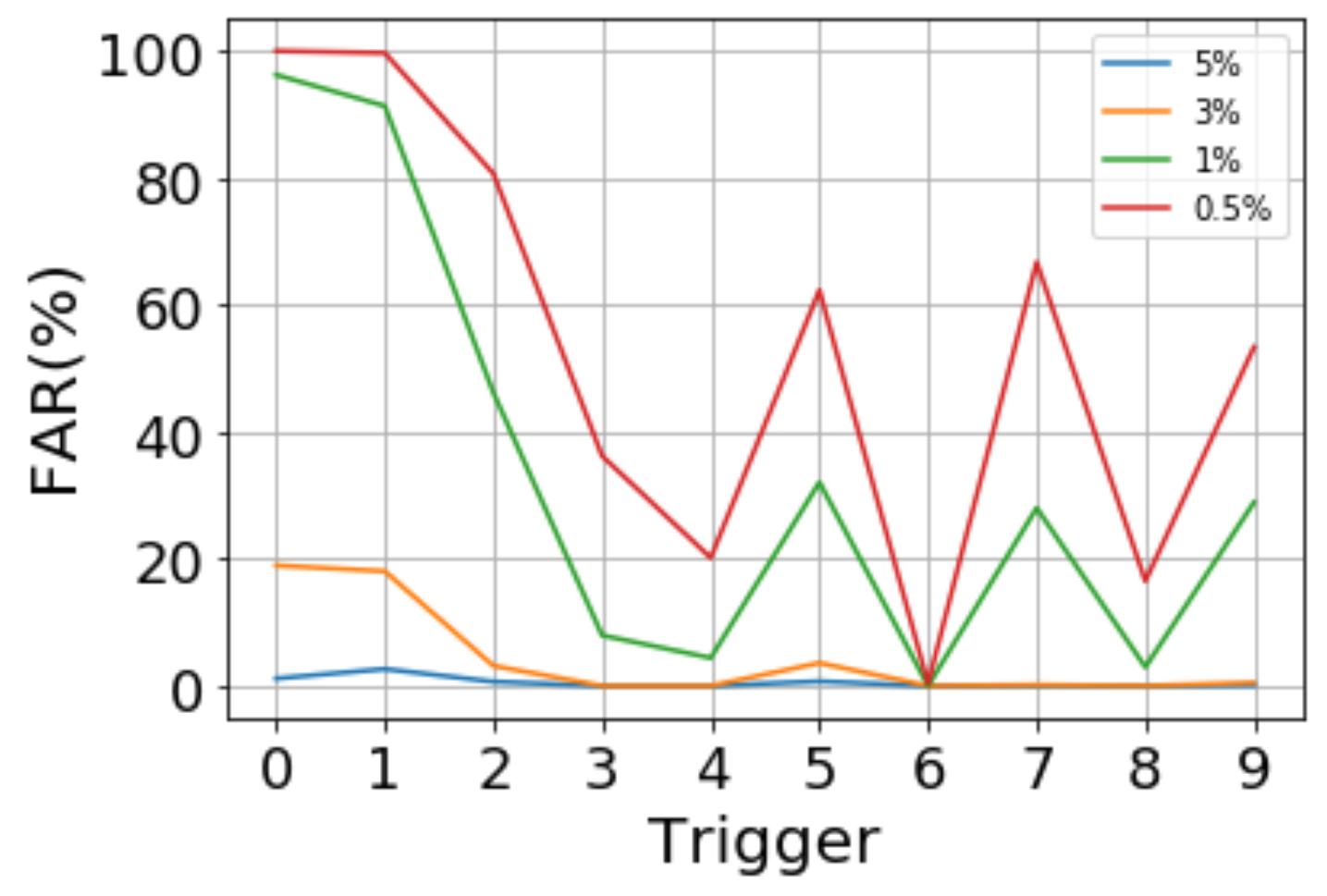}
	\caption{ Text task (CC+LSTM) detection capability for ten different triggers targeting ten different labels. The FRR settings of 0.5\%, 1\%, 3\% and 5\% are illustrated, respectively.}
	\label{fig:TextFAR_A2}
\end{figure}

\begin{figure}[h]
	\centering
	\includegraphics[trim=0 0 0 0,clip,width=0.5\textwidth]{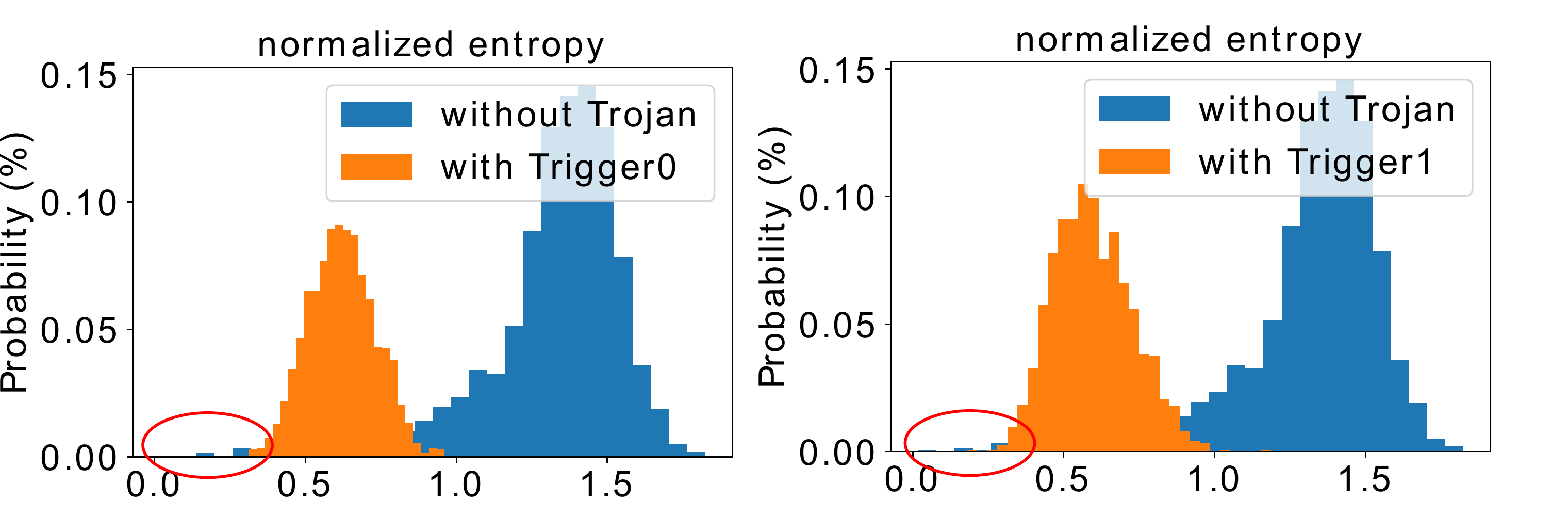}
	\caption{ Red circled region shows the anomaly entropy from clean inputs, and explains the extremely high FAR for e.g., trigger0 and trigger1, in Fig.\ref{fig:TextFAR_A2}.}
	\label{fig:Trigger_A2}
\end{figure}

From trigger 0 to 9, attack success rates are 99.22\%, 99.48\%, 99.36\%, 99.99\%, 98.79\%, 96.53\%, 100\%, 97.15\%, 98.86\%, 99.99\%, respectively. The detection capability for each trigger is detailed in Fig.~\ref{fig:TextFAR_A2}. It is noted that for e.g., trigger0, trigger1, the FAR is substantially high when the FRR is set to be too small, e.g., 0.5\% or 1\%. We investigate the entropy distribution in these cases. The reason is that there are some anomaly clean input accidentally exhibiting extremely low entropy, as shown in Fig.~\ref{fig:Trigger_A2}\footnote{Note that we use 2,000 samples for testing Trojan inputs in our experiments. Increasing the testing samples to determine the detection boundary can eliminate this issue as well since those clean inputs exhibiting extremely low entropy are rare.}. Nonetheless, we can see by using a slightly higher FRR, e.g., 5\%, the FAR can always be successfully suppressed to be lower than 3\% for all triggers. 

\subsection{Same Infected Label with Separate Triggers (A3)}

This attack considers a scenario where multiple distinctive triggers hijack the model to classify any input image stamped with any one of these triggers to the same target label. 

\subsubsection{Audio} 
\begin{figure}[h]
	\centering
	\includegraphics[trim=0 0 0 0,clip,width=0.5\textwidth]{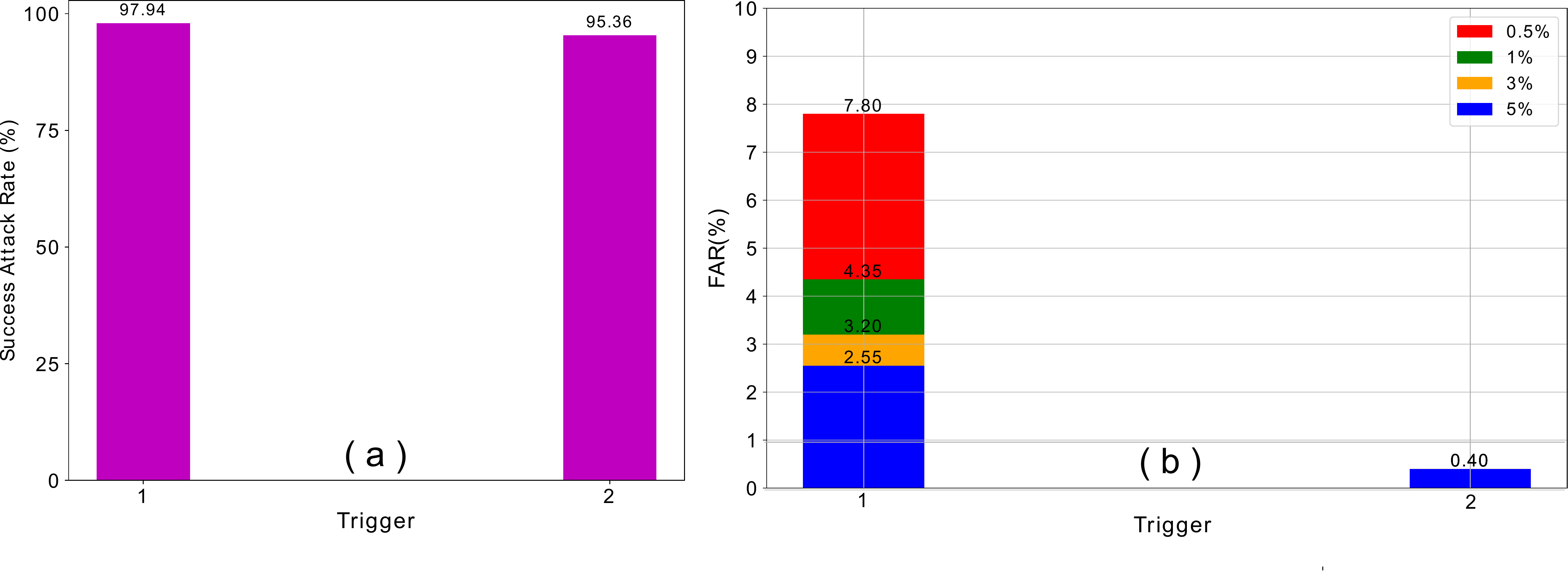}
	\caption{ (a) Attack success rate of different triggers targetting the same label. (b) Audio task detection capability for different triggers, where the FRR settings of 0.5\%, 1\%, 3\% and 5\% are illustrated, respectively.}
	\label{fig:AudioSeperateTriggerSameLabel}
\end{figure}

The Trojan attack settings are same to that in Section~\ref{sec:AudioSeperateTriggerDifferentLabel} except that now two triggers target the same label 'Zero'. The attack success rate and STRIP-ViTA detection performance are detailed in Fig.~\ref{fig:AudioSeperateTriggerSameLabel} (a) and (b) respectively. We can see that STRIP-ViTA is able to efficiently detect Trojan inputs stamped with any of these triggers during run-time. 

Notably, we found that the FAR of trigger2 keeps almost constant for all four FRR settings: 0.5\%, 1\%, 3\%, 5\%. The reason here is that out of 2000 tested Trojan inputs, only 5 preserve their Trojan effects. Recall here the FAR has excluded those Trojan inputs that cannot preserve their Trojan effects, as shown in the last column of Table~\ref{tab:FRRFARAudio}. We investigated the raw FAR when including those Trojan inputs lose their Trojaning effect, it is 17.65\%, 13.10\%, 7.95\%, 6.95\% corresponding to the FRR setting of 0.5\%, 1\%, 3\% and 5\%, respectively. We can see that for this trigger2, almost all of the Trojan inputs eventually do not have Trojan effects---only 5 preserve their Trojan effects always exhibiting higher entropy.

\subsection{Text}
The Trojan attack settings are same to that in Section~\ref{sec:TextSeperateTriggerDifferentLabel} except that now all ten triggers target the same label---the 4$_{\rm th}$ label. The Trojaned model classification rate for clean inputs is 64\%, lower than the clean model. For trigger 0 to 9, attack success rates are 99.45\%, 99.98\%, 100\%, 99.21\%, 99.82\%, 99.87\%, 99.45\%, 99.83\%, 99.88\%, and 99.94\%, respectively. 

\begin{figure}[h]
	\centering
	\includegraphics[trim=0 0 0 0,clip,width=0.3\textwidth]{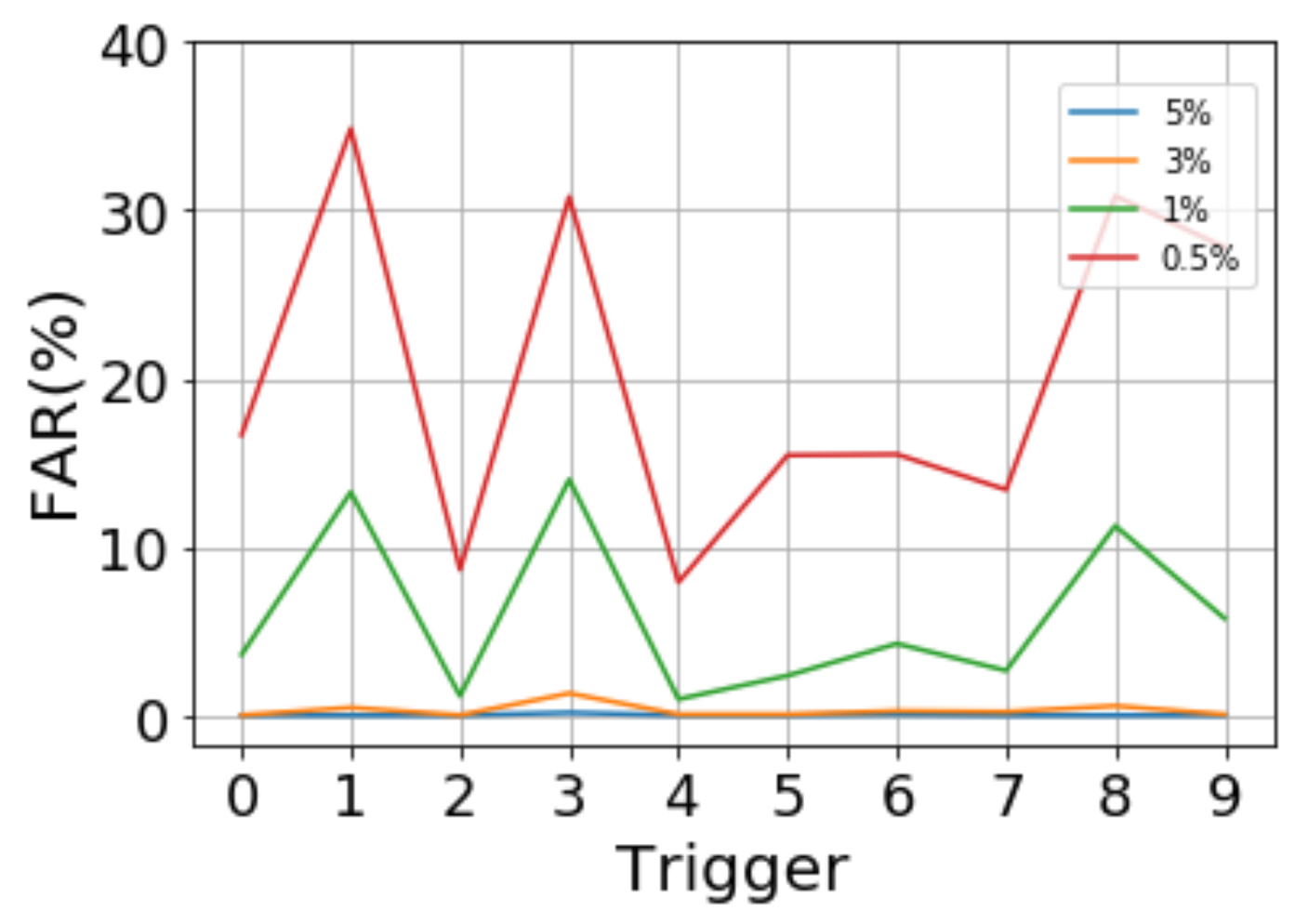}
	\caption{ Text task detection capability for ten different triggers targeting the same label. The FRR settings of 0.5\%, 1\%, 3\% and 5\% are illustrated, respectively.}
	\label{fig:TextFAR_A3}
\end{figure}

The detection capability for each trigger is detailed in Fig.~\ref{fig:TextFAR_A3}. We can see that when the FRR is set to be 3\%, FAR for each trigger is always less than 2\% for any trigger. 

\vspace{0.5cm}
\noindent{\bf Summery:} To this end, we can conclude that our STRIP-ViTA is efficiently against all the above advanced Trojan attacks. The crucial reason is that STRIP-ViTA exploits the input-agnostic Trojan characteristic---the main strength of such attack, while it is independent on trigger shape, size or/and other settings. In other words, as long as the trigger is input-agnostic and a high attack success rate is desired---always the case for the attacker, STRIP-ViTA will be effective regardless Trojan attack variants.

\section{Related Work and Comparison}\label{sec:RelatedWork}
We present related work including both Trojan attacks and the state-of-the-art countermeasures. We then compare STRIP-ViTA with these related countermeasures.

\subsection{Trojan Attacks}
In 2017, Gu {\it et al.}~\cite{gu2017badnets,gu2019badnets} proposed Badnets, where the attacker has access to the training data and can, thus, manipulate the training data via stamping arbitrarily chosen triggers---e.g., square-like trigger located at the corner of the digit image of the MNIST data---and then change the class labels. By training on those poisoned samples, a Trojaned model can be easily obtained. On the MNIST dataset, Gu {\it et al.}~\cite{gu2017badnets,gu2019badnets} demonstrated an attack success rate of over 99\% without impacting model performance on benign inputs. In addition, Trojan triggers to misdirect traffic sign classifications have also been investigated in~\cite{gu2017badnets}. Chen {\it et al.}~\cite{chen2017targeted} from UC Berkeley concurrently demonstrated such Trojan attacks by poisoning the training dataset. 

Liu~{\it et al.}~\cite{liu2018trojaning} eschewed the requirements of accessing the training data. Instead, their attack is performed during the model update phase, not model training phase. They first carry out reverse engineer to synthesize the training data, then improve the trigger generation process by delicately designing triggers to maximize the activation of chosen internal neurons in the neural network. This builds a stronger connection between triggers and internal neurons, thus, requiring less training samples to insert backdoors. A related work by Ji~{\it et al.}~\cite{ji2018model} inserted insidious malicious behaviour to a feature extractor---pretrained model. If victim trains their model by using the malicious feature extractor, their model will be infected.

Bagdasaryan~{\it et al.}~\cite{bagdasaryan2018backdoor} showed that federated learning is fundamentally vulnerable to trojan attacks. Firstly, participants are enormous, e.g., millions, it is impossible to guarantee that none of them are malicious. Secondly, federated learning is designed to have no access to the participant's local data and training process to ensure the privacy of the sensitive training data; therefore, participants can use Trojaned data for training. The authors demonstrate that with control over no more than 1\% participants, an attacker is able to cause a global model to be Trojaned and achieves a 100\% accuracy on the Trojaned input even when the attacker is only selected in a single round of training---federated learning requires a number of rounds to update the global model parameters. 
\subsection{Trojan Defences}
Trojan defences can be generally classified into two categories. The first is to inspect the model to determine whether the model itself is Trojaned or not---performed offline. The second is to check the input during run-time when the model is under deployment. If the Trojan input is detected, an rejection can be applied and then an alert can be thrown to allow the user to carefully inspect the input and potentially discover the trigger and further carry out model fixing.

\subsubsection{Offline Trojan Model Detection and Fix}
Works in~\cite{liu2018fine,liu2017neural} suggested approaches to remove the Trojan behavior without first checking whether the model is Trojaned or not. Fine-tuning is used to remove potential Trojans by pruning carefully chosen parameters of the DNN model~\cite{liu2018fine}. However, this method substantially degraded the model accuracy~\cite{wangneural}. Approaches presented in~\cite{liu2017neural} incurred high complexity and computation costs. 

In S\&P 2019, Wang {\it et al.}~\cite{wangneural} proposed the Neural Cleanse method to detect whether a DNN model has been Trojaned or not offline, where its accuracy was further improved in~\cite{guo2019tabor}. Neural Cleanse is based on the intuition that, given a Trojaned model, it requires much smaller modifications to all input samples to misclassify them into the attacker targeted (infected) label than any other uninfected labels. Therefore, their method iterates through all labels of the model and determines if any label requires a substantially smaller amount of modification to achieve misclassifications.
One advantage of this method is that the trigger can be discovered and identified during the Trojaned model detection process. Note that the reversed trigger may not have same or even similar visualisation to the original trigger. However, this method has two limitations. Firstly, it could incur high computation costs proportionally to the number of labels. 
Secondly, similar to SentiNet~\cite{chou2018sentinet}, the method is with decreasing effectiveness with increasing trigger size.


In CCS 2019, Liu {\it et al.} proposed Artificial Brain Stimulation (ABS)~\cite{liu2019abs} by scanning a DNN model to determine whether it is Trojaned. Inspiring from the Electrical Brain Stimulation (EBS) technique used to analyze the human brain neurons, Liu {\it et al.} created their ABS to inspect individual neuron activation difference for anomaly detection of Trojan. Some advantages of ABS are that i) it is trigger size independent, and ii) requires only one input per label to detect the Trojan. iii) It can also detect Trojan attacks on feature space rather besides pixel space. Nevertheless, the method appears to only effective under certain critical assumptions, e.g., the target label output activation needs to be activated by {\it only one} neuron instead of from interaction of a group of neurons. In addition, the scope is also limited to the attack of one single trigger per label. If multiple triggers were aimed to attack the same label, it would be out of ABS's reach. 

In IJCAI 2019, Chen {\it et al.} proposed DeepInspect~\cite{chen2019deepinspect} to detect Trojan attack even without any access to training data. The key idea of DeepInspect is to use a conditional generative model to learn the probabilistic distribution of potential triggers. This generative model will be used to generate reversed triggers whose their perturbation level will be statistically evaluated to build the Trojan anomaly detection. DeepInspect is faster than Neural Cleanse to reverse triggers in complex datasets. However, due to the strict assumption of having no access to training data, the result of this DeepInspect appears to be worse than other state-of-the-art methods in some situations.

\subsubsection{Run-time Trojan Input Detection}
Chou {\it et al.}~\cite{chou2018sentinet} exploited both the model interpretability and object detection techniques, referred to as SentiNet, to firstly discover contiguous regions of an input image important for determining the classification result. This region is assumed having a high chance of possessing a trojan trigger when it strongly affects the classification. Once this region is determined, it is carved out and patched on to other held-out images that are with ground-truth labels. If both the misclassification rate---probability of the predicted label is not the ground-truth label of the held-out image---and confidence of these patched images are high enough, this carved patch is regarded as an adversarial patch that contains a Trojan trigger. Therefore, the incoming input is a Trojaned input.

In NDSS 2019, Ma {\it et al.} proposed NIC~\cite{ma2019nic} by checking the provenance channel and activation value distribution channel. They extract DNN invariants and use them to perform run-time adversarial sample detection including trojan input detection. This method can be generally viewed to check the activation distribution and flow across DNN layers---inspired by the control flow used in programming---to determine whether the flow is violated due to the adversarial samples. However, NIC indeed requires extensive offline training to gain different classifiers across layers to check the activation distribution and flow and can be easily evaded by adaptive attacks~\cite{ma2019nic}.

Doan {\it et al.} proposed Februus~\cite{februus} to detect and sanitize the Trojaned input at run-time before it was fed into a Trojaned classifier. They utilized a Visual Explanation tool to detect the trigger region before surgically removing it. To maintain the performance of the network, a GAN-based image inpainting method has been deployed to restore the likeliness of the Trojan images before it was Trojaned. This method can be attached in front of any deep neural networks acting as a Trojan filter to eliminate the Trojan effect on the network. 



\begin{table*}
	\centering 
	\caption{Comparison with other trojan detection works.}
			\resizebox{0.85\textwidth}{!}{
	\begin{tabular}{c | c | c | c | c | c | c | c | c | c | c} %
		\toprule 
		\toprule 
				
		Work &  \begin{tabular}{@{}c@{}} Black/White  \\ -Box Access \end{tabular}  & \begin{tabular}{@{}c@{}} No Ground-Truth \\ Reference Model \\ Required \end{tabular} & \begin{tabular}{@{}c@{}} Run-time \\ Offline \end{tabular} & \begin{tabular}{@{}c@{}} No Trojaned  \\ Samples Required \end{tabular} & \begin{tabular}{@{}c@{}} Vision \\ (FRR, FAR) \end{tabular} & \begin{tabular}{@{}c@{}} Text \\ (FRR, FAR) \end{tabular} &  \begin{tabular}{@{}c@{}} Audio \\ (FRR, FAR) \end{tabular} & A1$^2$ & A2$^2$ & A3$^2$ \\ 
		\midrule
		
		\begin{tabular}{@{}c@{}} Neural Cleanse~\cite{wangneural} \\ (S\&P 2019) \end{tabular} &  Black-box & \ding{55} & Offline &  \ding{52} & (1.63\%, 5\%) & N/A$^3$ & N/A & \ding{55} & \textcolor{black}{\ding{51}}{\small\textcolor{black}{\kern-0.5em\ding{55}}} & \textcolor{black}{\ding{51}}{\small\textcolor{black}{\kern-0.5em\ding{55}}} \\
		\hline
		
		\begin{tabular}{@{}c@{}} DeepInspect~\cite{chen2019deepinspect} \\ (IJCAI 2019) \end{tabular} &  Black-box & \ding{52} & Offline &  \ding{52} & (0\%,0\%)$^4$ & N/A & N/A & \ding{55} & N/A & N/A \\ \hline
		
		\begin{tabular}{@{}c@{}} NIC~\cite{ma2019nic} \\ (NDSS 2019) \end{tabular} &  White-box & \ding{52} & Run-time &  \ding{52} & (3.4\%,0\%)$^1$ & N/A & N/A & N/A & N/A & N/A \\ \hline
		
		
		\begin{tabular}{@{}c@{}} ABS~\cite{chen2019deepinspect} \\ (CCS 2019) \end{tabular} &  White-box & \ding{55} & Offline &  \ding{52} & (N/A,1\%)$^5$ & N/A & N/A & \ding{52} & \ding{55} & \ding{55} \\ \hline		
		
	    \begin{tabular}{@{}c@{}} STRIP-ViTA \\ (Our Work) \end{tabular} &  Black-box & \ding{52} & Run-time &  \ding{52} & (0.125\%, 0\%) & (3\%, 1.1\%) & (3\%, 3.55\%) & \ding{52} & \ding{52} & \ding{52} \\ \hline	
		\bottomrule
	\end{tabular}}
	  \begin{tablenotes}
      \small
      \item $^1$ Result is from MNIST dataset. For ImageNet, to achieve 0\% FAR, the FRR needs to be up to 15.9\%.      
        
        \item $^2$ A1 (Trigger Size Independence), A2 (Multiple Infected Labels with Separate Triggers), A3 (Same Infected Label with Separate Triggers).
        \item $^3$ N/A means that results are not available.
        \item $^4$ DeepInspect determines whether the model is Trojaned or not. In~\cite{chen2019deepinspect}, in total, five different Trojaned models are evaluated.
        \item $^5$ The FAR is based on the average detection accuracy summerised in column 3 in Table 4 in~\cite{liu2019abs}. The FRR report is not available. 
    \end{tablenotes}
	\label{tab:comparison} 
\end{table*}

\vspace{-0.3cm}
\subsection{Comparison}\label{sec:limitation}
To this end, we compare STRIP-ViTA with other four recently published state-of-the-art defence works including Neural Cleanse~\cite{wangneural}, DeepInspect~\cite{chen2019deepinspect}, NIC~\cite{ma2019nic}, ABS~\cite{liu2019abs}, as summarised in Table~\ref{tab:comparison}. All of these works adopt the same assumption; having no access to Trojaned samples. There is a scenario where a defender can have access to the Trojaned samples~\cite{chen2018detecting,tran2018spectral} but we consider a common and weaker detection assumption same as Neural Cleanse, DeepInspect, NIC, and ABS.

\subsubsection{Across Domains} The STRIP-ViTA is the only detection that is validated across vision, text, and audio domains. All other detection methods are only validated on vision tasks, whether they are applicable to other domains remains unclear. Furthermore, we have also demonstrated that the STRIP-ViTA is independent on model architectures, e.g., i) LTSM and 1D CNN (Section~\ref{sec:Text1DCNN}), and ii) 1D CNN and 2D CNN (Section~\ref{sec:Audio2DCNN}), even for the same task by using the same dataset. While validations of all other countermeasures have been limited with 2D CNN for vision tasks. 

\subsubsection{Trojan Variants}
We have validated STRIP-ViTA efficiency against identified advanced backdoor attacks~\cite{wangneural} across {\it all three domain tasks}. Again, please kindly note that for all other works, they only evaluate their approaches against advanced backdoor variants based on {\it vision tasks}. 

Under vision task, only ABS and STRIP-ViTA are independent to trigger size (A1). Although Deepinspect is less sensitive to larger size triggers---the maximum trigger size reported ~\cite{chen2019deepinspect} is 25\% of the image input---in comparison to Neural Cleanse, it still appears not effective for larger size triggers. For A2 and A3, Neural Cleanse can only be effective on the condition that the number of infected labels, or the number of inserted triggers is small.

\subsubsection{Reference Model} Both Neural Cleanse and ABS require a ground-truth/golden reference model to determine the {\it global} detection boundary for distinguishing Trojaned model from clean model. Such global detection boundary might not be applicable to all Trojaned models in few cases. Probably, one not obvious fact is that users need to train Trojaned/clean models by themselves to find out this global setting relying on reference models. STRIP-ViTA does not need reference model but solely the already deployed (Trojaned/clean) model to determine the detection boundary that is unique to each deployed model. In addition, ground-truth reference model may partially violate the motivation for outsourcing the ML model training to third party---the main source of attackers to carry out backdoor attacks: if the users own training skills and the computational power, it may be reasonable to train the model, from scratch even without using pre-trained model---the other source of backdoor attack, by themselves.

Not obviously, one advantage of STRIP-ViTA is that it requires neither professional machine learning skills nor computational power to employ it---solely straightforward and simple entropy estimation. All other works either require training ground-truth reference model~\cite{wangneural,liu2019abs} or other classifiers~\cite{chen2019deepinspect,ma2019nic}. Again, if the user has such skills and computational resources, they may not opt for outsourcing to train their models that is one of the main source of Trojan attacks.

\subsubsection{Time Overhead} As different settings for different Trojan detection techniques (offline detection require no strict time for inspecting the model, though it is desirable to be faster), it is hard to fairly compare time overhead among them. For example, Neural Cleanse may take days, e.g., 17 days, depending on the model architecture and dataset. Here we opt for reporting time overhead of our STRIP-ViTA solely as a reference. 

For all above experimental tests, we set $N=100$ for consistence, however, this may not be the optimal number given the model architecture and dataset. Recall $N$ is the number of perturbing patterns applied to {\it one} given input. In practice, smaller $N$, faster STRIP-ViTA run-time overhead. To determine an optimal $N$, we have developed a method in~\cite{gao2019strip}. In general, we observe the standard variation of entropy distribution of clean inputs as a function of $N$. We pick up the smallest $N$ where the standard variation starts not change too much (details are referred to Section 5.4 in~\cite{gao2019strip}). Based on the determined $N$, we have evaluated the STRIP-ViTA run-time overhead~\cite{gao2019strip} through traffic recognition task (GTSRB) using ResNet20 for vision task, it takes 1.32 times longer---without optimisation---than the original inference time. 

Similarly, for text tasks, we have accordingly determined $N=20$ for IMDB+LSTM and $N=20$ for CC+LTSM, respectively. For IMDB, the default inference time for an input is 11.77~ms, while the STRIP-ViTA takes 49.33~ms. Thus, it is 4.24 times longer. For CC, the default inference time for an input is 83.56~ms, while the STRIP-ViTA takes 117.8ms, which is about 1.41 times longer.

For audio tasks, we have determined $N=10$ for SC+1D CNN. The default inference time for an audio input is 1.58~ms, while the STRIP-ViTA takes 2.56~ms, which is about 1.62 times longer.

Therefore, STRIP-ViTA is even acceptable for real-time applications (e.g., traffic recognition and speech recognition) that also shows insensitivity to model complexity and dataset. Please note for all the above reported time overhead of STRIP-ViTA, no further dedicated optimisation has been applied.

\section{Conclusion}\label{sec:conclusion}
In this work, we have corroborated the first multi-domain Trojan detection, STRIP-ViTA, that is applicable to not only vision but also text and audio domain tasks. We have extensively validated the effectiveness and performance of STRIP-ViTA through various model architectures and datasets. In addition, STRIP-ViTA exhibits high efficacy against a number of advanced input-agnostic backdoor attack variants validated across all three domains. 

Although we have developed different perturbation methods for each domain task, other specific and efficient perturbation methods can be applicable and leave future works. In addition, although out of the scope of threat assumption that the common Trojan attack is input-agnostic, it is imperative to develop Trojan detection method that is efficient to detect class-specific Trojan attacks, which appear to be an open challenge.

\section{Acknowledge}
We acknowledge useful suggestions from Chang Xu.

\end{document}